\documentclass[12pt]{article}
\usepackage{a4wide,amsmath,amssymb,bm,graphicx}
\newcommand{\D}{\rlap{\hspace{0.2em}/}D}
\newcommand{\MS}{\ensuremath{\overline{\text{MS}}}}
\DeclareMathOperator{\Tr}{Tr}
\numberwithin{equation}{section}

\begin{document}

\begin{flushright}
\newlength{\mplength}
\settowidth{\mplength}{SFB/CPP-09-00}
\begin{minipage}{\mplength}
TTP09-21\\
SFB/CPP-09-58
\end{minipage}
\end{flushright}

\renewcommand{\thefootnote}{\fnsymbol{footnote}}

\begin{center}
\LARGE Introduction to effective field theories%
\footnote{Lectures at the Helmholtz International School-Workshop
``Calculations for modern and future colliders'',
Dubna, July 10--20, 2009.}\\[0.5em]
\Large 1. Heisenberg--Euler effective theory,\\[0.5em]
\Large decoupling of heavy flavours\\[1.5em]
\large Andrey Grozin\\[0.5em]
\large Budker Institute of Nuclear Physics, Novosibirsk, and\\
\large Institut f\"ur Theoretische Teilchenphysik, Universit\"at Karlsruhe
\end{center}
\vspace{1.5em}

\begin{abstract}
This is the first part of lectures about effective field theories.
Decoupling of heavy-particle loops is considered
(heavy leptons in QED, heavy quarks in QCD).
\end{abstract}

\tableofcontents
\renewcommand{\thefootnote}{\arabic{footnote}}
\setcounter{footnote}{0}

\section{Introduction}
\label{S:Intro}

We don't know \emph{all} physics up to \emph{infinitely high} energies
(or down to \emph{infinitely small} distances).
Therefore, \emph{all} our theories are effective low-energy
(or large-distance) theories
(except \emph{The Theory of Everything}, if such a thing exists).

There is a high energy scale $M$
(and a short distance scale $1/M$)
where an effective theory breaks down.
We want to describe light particles (with masses $m_i\ll M$)
and their interactions at low energies,
i.e., with characteristic momenta $p_i\ll M$
(or, equivalently, at large distances $\gg1/M$).
To this end, we construct an effective Lagrangian
containing the light fields.
Physics at small distances $\lesssim1/M$ produces
local interactions of these fields.
The Lagrangian contains all possible operators
(allowed by symmetries of our theory).
Coefficients of operators of dimension $n+4$
are proportional to $1/M^n$.
If $M$ is much larger than energies we are interested in,
we can retain only renormalizable terms (dimension 4),
and, perhaps, a few power corrections.

These lectures can be used as an addendum to a standard
textbook on quantum field theory.
We follow notation of~\cite{G:07};
details of some computations omitted here can be found in this book.
Heisenberg--Euler Lagrangian is discussed in any QED textbook,
see also the review~\cite{D:04}.
Technically, the present lectures are largely devoted to QED.
But the same ideas and methods are also used in QCD,
only calculations are more lengthy:
more diagrams, colour factors\dots{}
The most clear treatment of decoupling by matching on-shell matrix elements
in the full theory and the effective theory
(where all loop corrections vanish because they contain no scale)
is presented in~\cite{CKS:98}.
The method of regions, which is a useful alternative
to constructing effective Lagrangians,
is discussed in the textbook~\cite{S:02}.

\section{Heisenberg--Euler effective theory}
\label{S:HE}

\subsection{Photonia}
\label{S:Photonia}

The first historical example of an effective low-energy theory
is the Heisenberg--Euler effective theory in QED.
It is still the best example
illustrating typical features of such theories.

\begin{figure}[ht]
\begin{center}
\begin{picture}(94,46)
\put(21,25){\makebox(0,0){\includegraphics{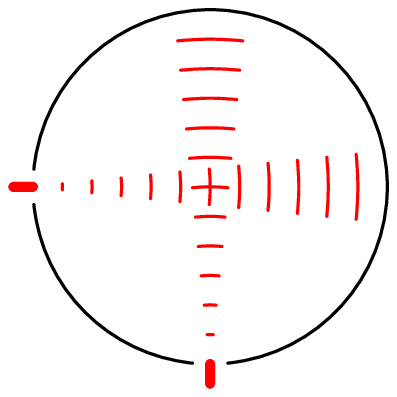}}}
\put(73.5,25){\makebox(0,0){\includegraphics{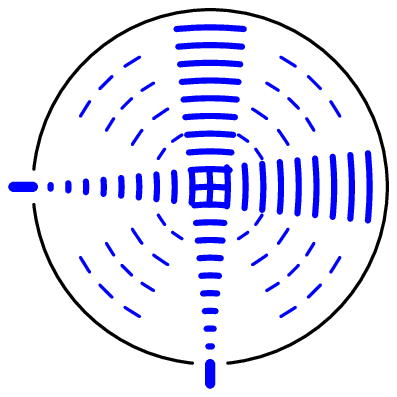}}}
\put(21,0){\makebox(0,0)[b]{a}}
\put(73,0){\makebox(0,0)[b]{b}}
\end{picture}
\end{center}
\caption{Scattering of low-energy photons}
\label{F:Photonia}
\end{figure}

In order to understand it better,
let's imagine a country, Photonia,
in which physicists have high-intensity sources
and excellent detectors of low-energy photons,
but they don't have electrons
and don't know that such a particle exists%
\footnote{We indignantly refuse to discuss the question
``What the experimentalists and their apparata are made of?''
as irrelevant.}.
At first their experiments (Fig.~\ref{F:Photonia}a)
show that photons do not interact with each other.
They construct a theory, Quantum PhotoDynamics (QPD),
with the Lagrangian
\begin{equation}
L_0 = - \frac{1}{4} F_{\mu\nu} F^{\mu\nu}\,.
\label{Photonia:L0}
\end{equation}
But later, after they increased the luminosity (and energy)
of their ``photon colliders'' and the sensitivity of their detectors,
they discover that photons do scatter,
though with a very small cross-section (Fig.~\ref{F:Photonia}b).
They need to add some interaction terms to this Lagrangian.

There are no dimension 6 gauge-invariant operators, because
\begin{equation}
F_\lambda{}^\mu F_\mu{}^\nu F_\nu{}^\lambda = 0
\label{Photonia:O6}
\end{equation}
(this algebraic fact reflects $C$-parity conservation).
The operator $F_{\mu\nu} \partial^2 F^{\mu\nu}$ can be rewritten as
\begin{equation}
F_{\mu\nu} \partial^2 F^{\mu\nu}
= - 2 F_{\mu\nu} \partial_\lambda \partial^\mu F^{\nu\lambda}
= 2 \partial_\mu \left( F_{\mu\nu} \partial_\lambda F^{\lambda\nu} \right)
- 2 \left( \partial^\mu F_{\nu\mu} \right)
\left( \partial_\lambda F^{\nu\lambda} \right)\,,
\label{Photonia:O61}
\end{equation}
where the identity
\begin{equation}
\partial_\lambda F_{\mu\nu} + \partial_\mu F_{\nu\lambda}
+ \partial_\nu F_{\lambda\mu} = 0
\label{Photonia:Bianci}
\end{equation}
has been used.
The full derivative can be omitted from the Lagrangian.
So, only one dimension 6 operator with two $F_{\mu\nu}$ remains,
\begin{equation}
O = \left( \partial^\mu F_{\lambda\mu} \right)
\left( \partial_\nu F^{\lambda\nu} \right)\,.
\label{Photonia:O62}
\end{equation}
But this operator vanishes due to equations of motion
\begin{equation}
\partial_\nu F^{\lambda\nu} = j^\lambda = 0\,.
\label{Photonia:EOM}
\end{equation}
On-shell matrix elements of such operators vanish;
therefore, we may omit them from the Lagrangian
without affecting the $S$-matrix.

Interaction operators first appear at dimension 8:
\begin{equation}
O_1 = \left(F_{\mu\nu} F^{\mu\nu}\right)^2\,,\qquad
O_2 = F_{\mu\nu} F^{\nu\alpha} F_{\alpha\beta} F^{\beta\mu}\,.
\label{Photonia:O8}
\end{equation}
Hence the QPD Lagrangian which incorporated the photon--photon interaction is
\begin{equation}
L = L_0 + L_1\,,\qquad
L_1 = c_1 O_1 + c_2 O_2\,,
\label{Photonia:L1}
\end{equation}
where the coefficients $c_{1,2}\sim1/M^4$,
$M$ is some large mass (the scale of new physics).
Of course, operators of dimensions $>8$ can be also included,
multiplied by higher powers of $1/M$,
but their effect at low energies is much smaller.
Physicists from Photonia can extract the two parameters $c_{1,2}$
from two experimental results,
and predict results of infinitely many measurements.

We are working at the order $1/M^4$;
therefore, the photon--photon interaction vertex
can appear in any diagram at most once.
The only photon self-energy diagram vanishes:
\begin{equation}
\raisebox{-12mm}{\includegraphics{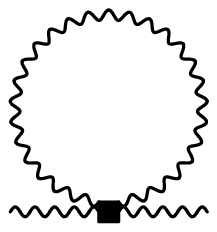}} = 0\,,
\label{Photonia:Pi0}
\end{equation}
because the massless loop is scale-free.
Therefore, the full photon propagator is equal to the free one.
There are no loop corrections to the 4-photon vertex,
and hence the interaction operators~(\ref{Photonia:O8})
don't renormalize.

\subsection{Qedland}
\label{S:Qedland}

In the neighboring country Qedland physicists are more advanced.
In addition to photons, they know electrons and positrons,
and investigate their interactions at energies $E\sim M$
($M$ is the electron mass).
They have constructed a nice theory, QED, which describe
their experimental results%
\footnote{They don't know muons, but this is another story
(Sect.~\ref{S:mu}).}.

Physicists from Qedland understand that QPD constructed in Photonia
is just a low-energy approximation to QED.
The coefficients $c_{1,2}$ can be calculated by matching:
we calculate the amplitude of photon--photon scattering
at low energies in both the full theory (QED)
and in the effective theory (QPD),
and equate them.
In QPD the scattering amplitude
is $T^{\mu_1\mu_2\nu_1\nu_2}(p_1,p_2,p'_1,p'_2)
e_{1\mu_1} e_{2\mu_2} e^{\prime*}_{1\nu_1} e^{\prime*}_{2\nu_2}$
where $T^{\mu_1\mu_2\nu_1\nu_2}=c_1 T_1^{\mu_1\mu_2\nu_1\nu_2}+c_2 T_2^{\mu_1\mu_2\nu_1\nu_2}$,
and the tensors $T_{1,2}^{\mu_1\mu_2\nu_1\nu_2}$
can be trivially obtained from the operators $O_{1,2}$~(\ref{Photonia:O8})
(they are linear in each momentum).

\begin{figure}[ht]
\begin{center}
\begin{picture}(150,46)
\put(23,23){\makebox(0,0){\includegraphics{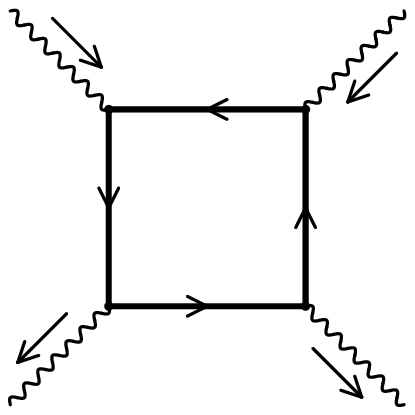}}}
\put(1,45){\makebox(0,0){$\mu_1$}}
\put(45,45){\makebox(0,0){$\mu_2$}}
\put(45,1){\makebox(0,0){$\nu_1$}}
\put(1,1){\makebox(0,0){$\nu_2$}}
\put(23,34.5){\makebox(0,0)[b]{$k$}}
\put(23,11.5){\makebox(0,0)[t]{$k+p_1-p_2$}}
\put(11.5,23){\makebox(0,0)[r]{$k+p_1$}}
\put(34.5,23){\makebox(0,0)[l]{$k-p_2$}}
\put(42,34){\makebox(0,0){$p_2$}}
\put(34,4){\makebox(0,0){$p_1$}}
\put(4,12){\makebox(0,0){$p_2$}}
\put(12,42){\makebox(0,0){$p_1$}}
\put(75,23){\makebox(0,0){\includegraphics{phia2a.eps}}}
\put(53,45){\makebox(0,0){$\mu_1$}}
\put(97,45){\makebox(0,0){$\mu_2$}}
\put(97,1){\makebox(0,0){$\nu_2$}}
\put(53,1){\makebox(0,0){$\nu_1$}}
\put(75,34.5){\makebox(0,0)[b]{$k$}}
\put(75,11.5){\makebox(0,0)[t]{$k$}}
\put(63.5,23){\makebox(0,0)[r]{$k+p_1$}}
\put(86.5,23){\makebox(0,0)[l]{$k-p_2$}}
\put(94,34){\makebox(0,0){$p_2$}}
\put(86,4){\makebox(0,0){$p_2$}}
\put(56,12){\makebox(0,0){$p_1$}}
\put(64,42){\makebox(0,0){$p_1$}}
\put(127,23){\makebox(0,0){\includegraphics{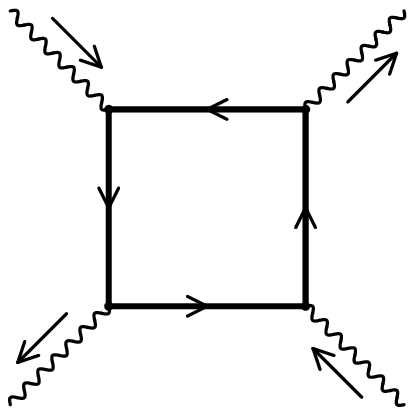}}}
\put(105,45){\makebox(0,0){$\mu_1$}}
\put(149,45){\makebox(0,0){$\nu_2$}}
\put(149,1){\makebox(0,0){$\mu_2$}}
\put(105,1){\makebox(0,0){$\nu_1$}}
\put(127,34.5){\makebox(0,0)[b]{$k$}}
\put(127,11.5){\makebox(0,0)[t]{$k$}}
\put(115.5,23){\makebox(0,0)[r]{$k+p_1$}}
\put(138.5,23){\makebox(0,0)[l]{$k+p_2$}}
\put(146,34){\makebox(0,0){$p_2$}}
\put(138,4){\makebox(0,0){$p_2$}}
\put(108,12){\makebox(0,0){$p_1$}}
\put(116,42){\makebox(0,0){$p_1$}}
\end{picture}
\end{center}
\caption{Photon--photon scattering in QED at one loop}
\label{F:gg1}
\end{figure}

In order to find 2 coefficients $c_{1,2}$,
it is sufficient to match just two scattering amplitudes.
For example, we can consider forward scattering:
$p'_1=p_1$, $p'_2=p_2$, $p_1^2=p_2^2=0$, $(p_1+p_2)^2=s\ll M^2$,
where $M$ is the electron mass.
It is sufficient to calculate 2 numbers in QED:
$T^{\mu_1\mu_2\nu_1\nu_2} g_{\mu_1\nu_1} g_{\mu_2\nu_2}$
and $T^{\mu_1\mu_2\nu_1\nu_2} g_{\mu_1\mu_2} g_{\nu_1\nu_2}$,
and then $c_{1,2}$ can be obtained by solving the linear system.
In QED, photon--photon scattering first appears at one loop
and is given by 6 diagrams;
pairs of diagrams which differ from each other
only by the direction of the electron line contribute equally
(just insert $U_C^2=1$ between all propagators and vertices,
where $U_C$ is the charge conjugation matrix:
$U_C \gamma_\mu U_C = - \gamma_\mu^T$, where $T$ means transposition).
Hence the scattering amplitude $T^{\mu_1\mu_2\nu_1\nu_2}(p_1,p_2,p_1,p_2)$
is twice the sum of 3 diagrams (Fig.~\ref{F:gg1}).

\begin{figure}[b]
\begin{center}
\begin{picture}(22,28.5)
\put(11,14){\makebox(0,0){\includegraphics{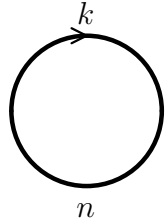}}}
\put(11,28.5){\makebox(0,0)[t]{$k$}}
\put(11,0){\makebox(0,0)[b]{$n$}}
\end{picture}
\end{center}
\caption{One-loop massive vacuum diagram}
\label{F:V1}
\end{figure}

In order to calculate the scalar quantities
$T^{\mu_1\mu_2\nu_1\nu_2} g_{\mu_1\nu_1} g_{\mu_2\nu_2}$
and $T^{\mu_1\mu_2\nu_1\nu_2} g_{\mu_1\mu_2} g_{\nu_1\nu_2}$,
we expand the propagators in $p_{1,2}\ll M$ up to the fourth order,
and average the integrand in $k$ directions.
The results are expressed via the one-loop vacuum integrals
(Fig.~\ref{F:V1})
\begin{equation}
\frac{1}{i \pi^{d/2}} \int \frac{d^d k}{D^n} = M^{d-2n} V(n)\,,\quad
D = M^2-k^2-i0\,,\quad
V(n) = \frac{\Gamma\left(n-\frac{d}{2}\right)}{\Gamma(n)}\,.
\label{Qedland:V1}
\end{equation}
Solving the linear system for $c_{1,2}$,
we arrive at the scattering amplitude
\begin{equation}
T^{\mu_1\mu_2\nu_1\nu_2} =
\frac{e_0^4 M^{-4-2\varepsilon}}{(4\pi)^{d/2}} \Gamma(\varepsilon)
\frac{(d-4)(d-6)}{2880}
\left( - 5 T_1^{\mu_1\mu_2\nu_1\nu_2} + 14 T_2^{\mu_1\mu_2\nu_1\nu_2} \right)\,.
\label{Qedland:M}
\end{equation}
It is finite at $\varepsilon\to0$.
To reproduce this QED result,
the interaction term in the QPD Lagrangian~(\ref{Photonia:L1})
must be~\cite{HE:36}
\begin{equation}
L_1 = \frac{\pi \alpha^2}{180 M^4}
\left( - 5 O_1 + 14 O_2 \right)\,.
\label{Qedland:L1}
\end{equation}

\begin{figure}[ht]
\begin{center}
\begin{picture}(74,22)
\put(11,11){\makebox(0,0){\includegraphics{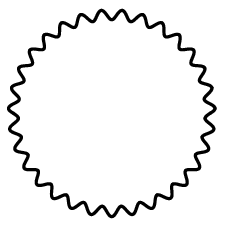}}}
\put(53,11){\makebox(0,0){\includegraphics{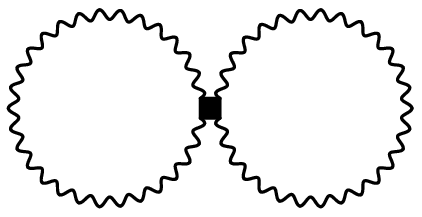}}}
\end{picture}
\end{center}
\caption{Energy density of thermal radiation}
\label{F:thermo}
\end{figure}

This effective Lagrangian can be applied to any problems
involving soft photons and their interactions.
As an example, let's consider the thermal radiation
at a temperature $T\ll M$.
What is its energy density?
The leading term is given by the first diagram
in Fig.~\ref{F:thermo}.
Note that this diagram is not scale-free and does not vanish,
because $T$ sets the scale.
The energy density is $\sim T^4$, by dimensionality
(Stefan--Boltzmann law).
The first correction is given by the second diagram
in Fig.~\ref{F:thermo}.
It contains the coupling $\sim\alpha^2/M^4$,
and, by dimensionality, it gives~\cite{KR:98}
\begin{equation*}
\sim \frac{\alpha^2}{M^4} T^8\,.
\end{equation*}
It would be difficult to guess this result
without using the effective theory.

\begin{figure}[b]
\begin{center}
\begin{picture}(22,29)
\put(11,14.5){\makebox(0,0){\includegraphics{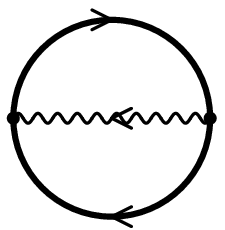}}}
\put(11,29){\makebox(0,0)[t]{$k_1$}}
\put(11,0){\makebox(0,0)[b]{$k_2$}}
\put(11,16){\makebox(0,0)[b]{$k_1-k_2$}}
\put(11,23){\makebox(0,0)[t]{$n_1$}}
\put(11,5.5){\makebox(0,0)[b]{$n_2$}}
\put(11,13){\makebox(0,0)[t]{$n_3$}}
\end{picture}
\end{center}
\caption{Two-loop massive vacuum diagram}
\label{F:V2}
\end{figure}

It is not difficult to calculate the two-loop correction
to the Lagrangian~(\ref{Qedland:L1}).
The two-loop scattering amplitude in QED
reduces to the two-loop vacuum integrals (Fig.~\ref{F:V2})
\begin{equation}
\begin{split}
&\frac{1}{(i\pi^{d/2})^2}
\int \frac{d^d k_1\,d^d k_2}{D_1^{n_1} D_2^{n_2} D_3^{n_3}}
= M^{2(d-n_1-n_2-n_3)} V(n_1,n_2,n_3)\,,\\
&D_1 = M^2 - k_1^2\,,\quad
D_2 = M^2 - k_2^2\,,\quad
D_3 = - (k_1-k_2)^2\,,\\
&V(n_1,n_2,n_3) =
\frac{\Gamma\left(\frac{d}{2}-n_3\right)
\Gamma\left(n_1+n_3-\frac{d}{2}\right)\Gamma\left(n_2+n_3-\frac{d}{2}\right)
\Gamma(n_1+n_2+n_3-d)}%
{\Gamma\left(\frac{d}{2}\right)\Gamma(n_1)\Gamma(n_2)\Gamma(n_1+n_2+2n_3-d)}
\end{split}
\label{Qedland:V2}
\end{equation}
($-i0$ is assumed in all denominators).

\subsection{Coulomb potential}
\label{S:Coulomb}

Let's suppose that physicists in Photonia have some classical
(infinitely heavy) charged particles,
and can manipulate them at their will.
If a particle with charge $e$ moves along a world line $l$,
the action contains the interaction term
\begin{equation}
S_{\text{int}} = e \int_l d x^\mu A_\mu(x)
\label{Coulomb:S}
\end{equation}
in addition to the photon field action.
The integrand $\exp(iS)$ in the Feynman path integral
contains a phase factor
\begin{equation}
W_l = \exp \left(i e \int_l d x^\mu A_\mu(x) \right)
\label{Coulomb:W}
\end{equation}
called the Wilson line (when the operator language is used,
the exponent is path-ordered).
The vacuum-to-vacuum transition amplitude in the presence
of classical charges is thus the vacuum average
of the corresponding Wilson lines%
\footnote{Later we'll see that the propagator of a heavy charged particle
in the effective theory which describes its interaction with soft photons
in the Wilson line.}.

\begin{figure}[ht]
\begin{center}
\begin{picture}(20,33)
\put(10,16.5){\makebox(0,0){\includegraphics{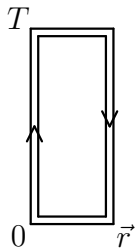}}}
\put(3,2){\makebox(0,0){$0$}}
\put(17,2){\makebox(0,0){$\vec{r}$}}
\put(3,31){\makebox(0,0){$T$}}
\end{picture}
\end{center}
\caption{Wilson loop}
\label{F:Wilson}
\end{figure}

Suppose two charges $e$ and $-e$ stay at rest
at some distance $\vec{r}$ during some (large) time $T$.
The energy of this system is $U(\vec{r}\,)$ ---
the interaction potential of the charges.
The vacuum transition amplitude is
\begin{equation}
e^{-i U(\vec{r}\,) T}\,.
\label{Coulomb:vac}
\end{equation}
On the other hand, this amplitude is the vacuum average
of the Wilson loop (Fig.~\ref{F:Wilson}) with $T\gg r$
(we don't care what happens near its lower and upper ends
as long as these regions are small as compared to $T$).

The zeroth-order term in the vacuum average of any Wilson loop is 1.
It is convenient (though not necessary) to use the Coulomb gauge
to calculate the first correction.
In this gauge, there is Coulomb photon with the propagator
\begin{equation}
D^{00}(q) = - \frac{1}{\vec{q}\,^2}
\label{Coulomb:D00}
\end{equation}
(it propagates instantaneously)
and transverse photon with the propagator
\begin{equation}
D^{ij}(q) = \frac{1}{q^2+i0}
\left(\delta^{ij}-\frac{q^i q^j}{\vec{q}\,^2}\right)\,.
\label{Coulomb:Dij}
\end{equation}
Wilson lines along the 0 direction only interact with Coulomb photons.
The self-energy of the classical particle vanishes:
\begin{equation}
\raisebox{-15.2mm}{\includegraphics{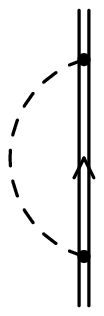}} = 0\,,
\label{Coulomb:self}
\end{equation}
because the particle propagates along time,
and the Coulomb photon along space.

Therefore, there is just one contribution at the order $e^2$:
\begin{equation}
\raisebox{-15.7mm}{\begin{picture}(24,33)
\put(10,16.5){\makebox(0,0){\includegraphics{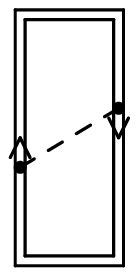}}}
\put(3,2){\makebox(0,0){$0$}}
\put(17,2){\makebox(0,0){$\vec{r}$}}
\put(3,31){\makebox(0,0){$T$}}
\put(2.5,13.5){\makebox(0,0){$\tau$}}
\put(21.5,19.5){\makebox(0,0){$\tau+t$}}
\end{picture}}
= - i\,e^2\,T\,\int D^{00}(t,\vec{r})\,dt
= - i\,e^2\,T\,\int \frac{d^{d-1}\vec{q}}{(2\pi)^{d-1}}\,
D^{00}(0,\vec{q})\,e^{i\,\vec{q}\cdot\vec{r}}
\label{Coulomb:W1}
\end{equation}
(integration in $\tau$ gives $T$).
Comparing it with $1 - i U(\vec{r}\,) T$~(\ref{Coulomb:vac}),
we obtain the Fourier transform of the potential
\begin{equation}
U(\vec{q}\,) = e^2 D^{00}(0,\vec{q}\,) = - \frac{e^2}{\vec{q}\,^2}\,;
\label{Coulomb:Uq}
\end{equation}
at $d=4$ the Coulomb potential is
\begin{equation}
U(\vec{r}\,) = - \frac{\alpha}{r}\,.
\label{Coulomb:U}
\end{equation}

\begin{figure}[ht]
\begin{center}
\begin{picture}(70,35)
\put(11,19){\makebox(0,0){\includegraphics{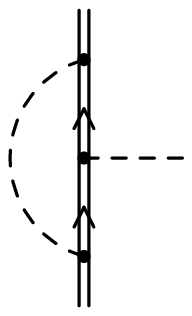}}}
\put(11,0){\makebox(0,0)[b]{$a$}}
\put(39,19){\makebox(0,0){\includegraphics{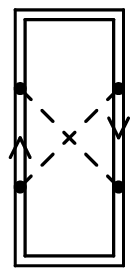}}}
\put(39,0){\makebox(0,0)[b]{$b$}}
\put(63,19){\makebox(0,0){\includegraphics{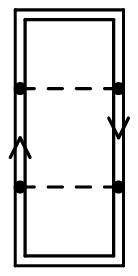}}}
\put(63,0){\makebox(0,0)[b]{$c$}}
\end{picture}
\end{center}
\caption{Corrections to the Wilson loop}
\label{F:Box}
\end{figure}

What about corrections?
Vertex corrections (Fig.~\ref{F:Box}$a$) vanish
for the same reason as~(\ref{Coulomb:self}).
Crossed-box diagrams like in Fig.~\ref{F:Box}$b$
vanish because Coulomb photons propagate instantaneously,
and the time orderings of the vertices on the left line
and on the right one cannot be opposite.
We don't need two-particle-reducible diagrams like in Fig.~\ref{F:Box}$c$
because they match higher orders of expansion
of the exponent~(\ref{Coulomb:vac}).
Only corrections to the photon propagator can contribute.
But there are no such corrections in QPD.
Hence the Coulomb potential~(\ref{Coulomb:U})
is exact in this theory.

In the presence of sources of the photon field,
the dimension 6 operator $O$~(\ref{Photonia:O62})
cannot be ignored.
The QPD Lagrangian now contains an extra term
\begin{equation}
L_c = c O\,,
\label{Coulomb:Lc}
\end{equation}
where $c\sim1/M^2$ by dimensionality.
This term produces the contribution to the photon self-energy
\begin{equation}
\raisebox{-5mm}{\begin{picture}(26,9.5)
\put(13,4.75){\makebox(0,0){\includegraphics{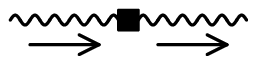}}}
\put(6.5,0){\makebox(0,0)[b]{$q$}}
\put(19.5,0){\makebox(0,0)[b]{$q$}}
\put(1,9.5){\makebox(0,0)[t]{$\mu$}}
\put(25,9.5){\makebox(0,0)[t]{$\nu$}}
\end{picture}}
= 2 i c q^2 \left(q^2 g_{\mu\nu} - q_\mu q_\nu\right)\,.
\label{Coulomb:Pic}
\end{equation}
The aim of an effective theory is to reproduce the $S$-matrix,
so, we may neglect operators vanishing due to equations of motion
(in particular, the self-energy~(\ref{Coulomb:Pic}) vanishes
at the photon mass shell $q^2=0$,
but is needed in a virtual photon line
exchanged between classical sources).
Therefore, the term~(\ref{Coulomb:Lc}) reduces to $c j_\mu j^\mu$,
where $j^\mu$ is the external (classical) current.
For the classical charges $e$ and $-e$,
it leads to the contact interaction%
\footnote{It is difficult to observe a $\delta$-function potential
in the interaction of classical charged particles.
But this interaction is essential if the particles
are quantum-mechanical --- it shifts energies of $S$-wave states
(we'll discuss this later, in NRQED).}
\begin{equation}
U_c(\vec{r}\,) = 2 c \delta(\vec{r}\,)\,.
\label{Coulomb:Uc}
\end{equation}

What can physicists in Qedland say about the interaction potential
between classical charged particles?
The photon self-energy in QED at one loop is
\begin{equation}
\raisebox{-11mm}{\begin{picture}(32,23)
\put(16,11.5){\makebox(0,0){\includegraphics{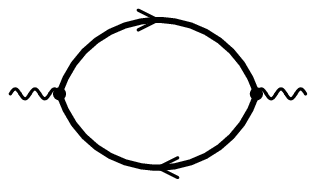}}}
\put(16,0){\makebox(0,0)[b]{$k$}}
\put(16,23){\makebox(0,0)[t]{$k+q$}}
\put(0,10.5){\makebox(0,0)[tl]{$\mu$}}
\put(32,10.5){\makebox(0,0)[tr]{$\nu$}}
\end{picture}} =
i \left( q^2 g_{\mu\nu} - q_\mu q_\nu \right) \Pi(q^2)\,.
\label{Coulomb:Pi}
\end{equation}
It is convenient to contract this with $g^{\mu\nu}$:
\begin{equation}
\begin{split}
&\Pi(q^2) = \frac{4 i e_0^2}{(d-1) q^2}
\int \frac{d^d k}{(2\pi)^d} \frac{N}{D_1 D_2}\,,\quad
D_1 = M_0^2 - (k+q)^2\,,\quad
D_2 = M_0^2 - k^2\,,\\
&N = \frac{1}{4} \Tr \gamma_\mu (\rlap/k+\rlap/q+M_0) \gamma^\mu (\rlap/k+M_0)
= \frac{d-2}{2} (D_1+D_2+q^2) + 2 M_0^2
\end{split}
\label{Coulomb:Pi1i}
\end{equation}
(here $e_0$ is the bare electron charge,
and $M_0$ is the bare electron mass).
When $q\ll M$, we can expand $1/D_1$ in $q$;
then averaging over $k$ directions reduces the problem
to the scalar integrals~(\ref{Qedland:V1}).
The result is
\begin{equation}
\Pi(q^2) = - \frac{4}{3} \frac{e_0^2 M_0^{-2\varepsilon}}{(4\pi)^{d/2}}
\Gamma(\varepsilon)
\left( 1 - \frac{d-4}{10} \frac{q^2}{M_0^2} + \cdots \right)\,.
\label{Coulomb:Pi1}
\end{equation}

If we retain only the leading term in~(\ref{Coulomb:Pi1}),
then the potential $U(\vec{q}\,)$ in QED
has the Coulomb form~(\ref{Coulomb:Uq}).
From now on, the charge in the low-energy effective theory (QPD)
will be denoted $e'$ (it is not renormalized: $e'=e_0'$).
The interaction potentials in both theories match if
\begin{equation}
e^{\prime2} = e_0^2 \left[ 1
- \frac{4}{3} \frac{e_0^2 M_0^{-2\varepsilon}}{(4\pi)^{d/2}} \Gamma(\varepsilon)
\right]\,.
\label{Coulomb:Match}
\end{equation}
The renormalized QED charge $e(\mu)$ in the \MS{} renormalization scheme
is related to the bare one $e_0$ by
\begin{equation}
e_0^2 = e^2(\mu) Z_\alpha(\alpha(\mu))\,,
\label{Coulomb:Renorm}
\end{equation}
where
\begin{equation}
Z_\alpha(\alpha) = 1 - \beta_0 \frac{\alpha}{4\pi\varepsilon} + \cdots
\label{Coulomb:Zalpha}
\end{equation}
and
\begin{equation}
\frac{\alpha(\mu)}{4\pi} =
\frac{e^2(\mu)\,\mu^{-2\varepsilon}}{(4\pi)^{d/2}}
e^{-\gamma_E \varepsilon}
\label{Coulomb:MSbar}
\end{equation}
($\gamma_E$ is the Euler constant).
Let's express the running QED charge $\alpha(\mu)$ via the QPD charge $e'$
from~(\ref{Coulomb:Match}) with the one-loop accuracy:
\begin{equation}
\frac{\alpha(\mu)}{4\pi} =
\frac{e^{\prime2} \mu^{-2\varepsilon}}{(4\pi)^{d/2}} e^{-\gamma_E \varepsilon}
\left[ 1 +
\frac{e^{\prime2} M^{-2\varepsilon}}{(4\pi)^{d/2}}
\left( \frac{\beta_0}{\varepsilon} \left(\frac{M}{\mu}\right)^{2\varepsilon}
+ \frac{4}{3} e^{\gamma_E \varepsilon} \Gamma(\varepsilon) \right)
+ \cdots \right]\,.
\label{Coulomb:alpha1}
\end{equation}
In the correction term we may replace $M_0$ by a renormalized
electron mass $M$
(in any scheme --- differences are beyond our current accuracy).
Both $e'$ and $\alpha(\mu)$ are finite at $\varepsilon\to0$.
Therefore,
\begin{equation}
\beta_0 = - \frac{4}{3}\,.
\label{Coulomb:beta0}
\end{equation}
The running QED coupling $\alpha(\mu)$
in the physical limit $\varepsilon\to0$
is expressed via the QPD $\alpha'=e^{\prime2}/(4\pi)$
(which is measured in macroscopic experiments) by
\begin{equation}
\alpha(\mu) = \alpha' \left( 1
- \beta_0 \frac{\alpha'}{4\pi} \log\frac{\mu^2}{M^2} + \cdots \right)\,.
\label{Coulomb:alpha}
\end{equation}
It satisfies the renormalization group equation
\begin{equation}
\frac{d\,\log\alpha(\mu)}{d\,\log\mu} = - 2 \beta(\alpha(\mu))\,,\quad
\beta(\alpha) = \beta_0 \frac{\alpha}{4\pi} + \cdots
\label{Coulomb:RG}
\end{equation}
(see~(\ref{Coulomb:beta0}))
with the initial condition
\begin{equation}
\alpha(M) = \alpha'\,.
\label{Coulomb:init}
\end{equation}

What about the $q^2$ term in~(\ref{Coulomb:Pi1})?
Comparing it with~(\ref{Coulomb:Pic}) we obtain
\begin{equation}
c = - \frac{2}{15} \frac{\alpha}{4\pi} \frac{1}{M^2}
\label{Coulomb:c6}
\end{equation}
at $\varepsilon=0$,
and thus the contact interaction~(\ref{Coulomb:Uc}).
We can also see this more directly:
this $q^2$ term cancels the denominator of the photon propagator
in~(\ref{Coulomb:Uq}), and thus give the contribution
\begin{equation}
U_c(\vec{q}\,) = - \frac{4}{15} \frac{\alpha^2}{M^2}\,,\quad
U_c(\vec{r}\,) = - \frac{4}{15} \frac{\alpha^2}{M^2} \delta(\vec{r}\,)
\label{Coulomb:Ucres}
\end{equation}
to the interaction potential at $\varepsilon=0$.

\subsection{The full theory (QED) and the low-energy effective theory (QPD)}
\label{S:Decoupling}

Now we shall discuss relations of these two theories
more systematically.
The Lagrangian of the low-energy effective theory (QPD) is
\begin{equation}
L' = - \frac{1}{4} F'_{0\mu\nu} F_0^{\prime\mu\nu}
- \frac{1}{2a'_0} \left(\partial_\mu A_0^{\prime\mu}\right)^2
\label{Decoupling:LQPD}
\end{equation}
(primed quantities are those of QPD;
we neglect the quartic interaction terms for now).
There are no loops which can be inserted into the photon propagator
in this theory, and hence no renormalization:
\begin{equation}
A'_0 = A'(\mu) = A'_{\text{os}}\,,\qquad
a'_0 = a'(\mu) = a'_{\text{os}}
\label{Decoupling:noren}
\end{equation}
(the index ``os'' means the on-shell renormalization scheme;
quantities with the argument $\mu$ are those in the \MS{} scheme).

The QED bare fields and parameters are related to renormalized ones
in the \MS{} scheme by
\begin{equation}
\begin{split}
&\psi_0 = Z_\psi^{1/2}(\alpha(\mu)) \psi(\mu)\,,\quad
A_0 = Z_A^{1/2}(\alpha(\mu)) A(\mu)\,,\\
&a_0 = Z_A(\alpha(\mu)) a(\mu)\,,\quad
e_0 = Z_\alpha^{1/2}(\alpha(\mu)) e(\mu)\,,\quad
M_0 = Z_m(\alpha(\mu)) M(\mu)\,,
\end{split}
\label{Decoupling:ren}
\end{equation}
where all renormalization constants have the minimal form
\begin{equation}
Z_i(\alpha) = 1 + \frac{z_1}{\varepsilon} \frac{\alpha}{4\pi}
+ \left(\frac{z_{22}}{\varepsilon^2} + \frac{z_{21}}{\varepsilon}\right)
\left(\frac{\alpha}{4\pi}\right)^2 + \cdots
\label{Decoupling:min}
\end{equation}
and $\alpha(\mu)$ is defined by~(\ref{Coulomb:MSbar}).

Another renormalization scheme often used in QED
is the on-shell scheme.
The photon field renormalized in this scheme
is related to the bare one by
\begin{equation}
A_0 = \left(Z_A^{\text{os}}\right)^{1/2} A_{\text{os}}
\label{Decoupling:Aos}
\end{equation}
(where the renormalization constant $Z_A^{\text{os}}$ is not minimal),
and therefore the bare and renormalized photon propagators
are related by
\begin{equation}
D_\bot(p^2) = Z_A^{\text{os}} D_\bot^{\text{os}}(p^2)\,.
\label{Decoupling:Dos}
\end{equation}
The photon propagator near the mass shell is
\begin{equation}
D_\bot(p^2) = \frac{1}{1 - \Pi(p^2)} \frac{1}{p^2}
= \frac{1}{1 - \Pi(0)} \frac{1}{p^2} + \cdots
\label{Decoupling:Dnms}
\end{equation}
By definition, $D_\bot^{\text{os}}(p^2)$ behaves
as the free propagator $1/p^2$ near the mass shell.
Therefore,
\begin{equation}
Z_A^{\text{os}} = \frac{1}{1 - \Pi(0)}\,.
\label{Decoupling:ZAos}
\end{equation}

We want to calculate processes with low-energy photons in QED,
and compare its predictions with those of the low-energy theory, QPD.
First of all, we shall consider the photon propagator.
The propagators of both $A_{\text{os}}$ and $A'_{\text{os}}$ at $p^2\to0$
are equal to the free propagator, and therefore
\begin{equation}
A_{\text{os}} = A'_{\text{os}}\,.
\label{Decoupling:Arel}
\end{equation}
The bare photon fields in the two theories are related by
\begin{equation}
A_0 = \left(\zeta_A^0\right)^{1/2} A'_0
\label{Decoupling:DecA0}
\end{equation}
(up to corrections suppressed by powers of $1/M$).
From~(\ref{Decoupling:Arel}) we obtain the bare decoupling coefficient
\begin{equation}
\zeta_A^0 = Z_A^{\text{os}}\,.
\label{Decoupling:zetaA0}
\end{equation}
The corresponding relation between the \MS{} renormalized fields is
\begin{equation}
A(\mu) = \zeta_A^{1/2}(\mu) A'(\mu)\,,
\label{Decoupling:ArelMS}
\end{equation}
where the renormalized decoupling coefficient is
\begin{equation}
\zeta_A(\mu) = \frac{\zeta_A^0}{Z_A} = \frac{Z_A^{\text{os}}}{Z_A}\,.
\label{Decoupling:zetaA}
\end{equation}

At one loop
\begin{equation}
\left(\zeta_A^0\right)^{-1} = \left(Z_A^{\text{os}}\right)^{-1} = 1 - \Pi(0)
= 1 + \frac{4}{3} \frac{e_0^2 M_0^{-2\varepsilon}}{(4\pi)^{d/2}}
\Gamma(\varepsilon) + \cdots
\label{Decoupling:zetaA01}
\end{equation}
(see~(\ref{Coulomb:Pi1}).
We can re-express it via renormalized quantities:
\begin{equation}
\frac{e_0^2 M_0^{-2\varepsilon}}{(4\pi)^{d/2}} \Gamma(\varepsilon) =
e^{L\varepsilon} e^{\gamma\varepsilon} \Gamma(1+\varepsilon)
\frac{\alpha(\mu)}{4\pi\varepsilon} Z_\alpha Z_m^{-2\varepsilon}\,,\qquad
L = 2 \log\frac{\mu}{M(\mu)}\,.
\label{Decoupling:via}
\end{equation}
The renormalized decoupling coefficient
$\zeta_A^{-1}=Z_A(\zeta_A^0)^{-1}=Z_A/Z_A^{\text{os}}$
must be finite at $\varepsilon\to0$ (for example, at $L=0$).
We substitute the \MS{} renormalization constant
$Z_A=1+z_1 \alpha(\mu)/(4\pi\varepsilon)$
with an unknown $z_1$, and find from this requirement
\begin{equation}
Z_A = 1 - \frac{4}{3} \frac{\alpha(\mu)}{4\pi\varepsilon} + \cdots
\label{Decoupling:ZA1}
\end{equation}
Finally, we arrive at
\begin{equation}
\zeta_A^{-1}(\mu) = 1 + \frac{4}{3} L \frac{\alpha(\mu)}{4\pi}
+ \cdots
\label{Decoupling:zetaA1}
\end{equation}

\begin{figure}[ht]
\begin{center}
\begin{picture}(106,17)
\put(16,8.5){\makebox(0,0){\includegraphics{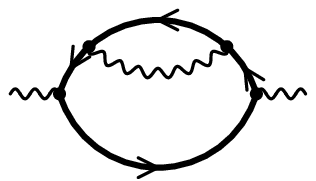}}}
\put(53,8.5){\makebox(0,0){\includegraphics{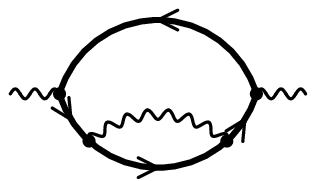}}}
\put(90,8.5){\makebox(0,0){\includegraphics{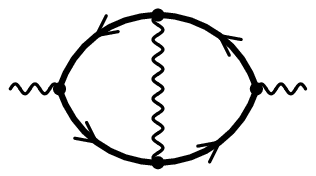}}}
\end{picture}
\end{center}
\caption{Two-loop photon self-energy}
\label{F:Photon2}
\end{figure}

Calculation of $\Pi(0)$ at two loops (Fig.~\ref{F:Photon2})
reduces to the vacuum integrals~(\ref{Qedland:V2}),
and it is not difficult to obtain the result
\begin{equation}
\Pi(0) = - \frac{4}{3} \frac{e_0^2 M_0^{-2\varepsilon}}{(4\pi)^{d/2}} \Gamma(\varepsilon)
- \frac{2}{3} \frac{(d-4)(5d^2-33d+34)}{d(d-5)}
\left(\frac{e_0^2 M_0^{-2\varepsilon}}{(4\pi)^{d/2}}
\Gamma(\varepsilon)\right)^2 + \cdots
\label{Decoupling:Pi2}
\end{equation}

Expressing the on-shell renormalization constant~(\ref{Decoupling:ZAos})
(and the bare decoupling coefficient~(\ref{Decoupling:zetaA0}))
via renormalized quantities~(\ref{Decoupling:via}) we obtain
\begin{equation}
\begin{split}
\left(\zeta_A^0\right)^{-1} = \left(Z_A^{\text{os}}\right)^{-1} ={}& 1 - \Pi(0)
= 1 + \frac{4}{3} e^{L\varepsilon} 
\frac{\alpha(\mu)}{4\pi\varepsilon} Z_\alpha Z_m^{-2\varepsilon}\\
&{} - \varepsilon \left(6 - \frac{13}{3} \varepsilon + \cdots\right)
e^{2L\varepsilon} \left(\frac{\alpha(\mu)}{4\pi\varepsilon}\right)^2
+ \cdots
\end{split}
\label{Decoupling:zA0}
\end{equation}
When re-expressing it via renormalized quantities,
we have to use one-loop renormalization constants
in the one-loop term.
The charge renormalization constant can be obtained
from the photon-field one~(\ref{Decoupling:ZA1}):
\begin{equation}
Z_\alpha = Z_A^{-1} = 1 + \frac{4}{3} \frac{\alpha(\mu)}{4\pi\varepsilon}
+ \cdots
\label{Decoupling:Zalpha1}
\end{equation}
The \MS{} mass renormalization constant is
\begin{equation}
Z_m = 1 - 3 \frac{\alpha(\mu)}{4\pi\varepsilon} + \cdots
\label{Decoupling:Zm1}
\end{equation}
(see~(\ref{App:gammam})).
The renormalized decoupling coefficient
$\zeta_A^{-1}=Z_A/Z_A^{\text{os}}$
must be finite at $\varepsilon\to0$ (for example, at $L=0$).
We substitute the \MS{} renormalization constant
$Z_A=1-(4/3)\alpha(\mu)/(4\pi\varepsilon)
+(z_{20}+z_{21}\varepsilon)(\alpha(\mu)/(4\pi\varepsilon))^2$
with unknown coefficients $z_{20}$, $z_{21}$,
and find from this requirement
\begin{equation}
Z_A = 1 - \frac{4}{3} \frac{\alpha(\mu)}{4\pi\varepsilon}
- 2 \varepsilon \left(\frac{\alpha(\mu)}{4\pi\varepsilon}\right)^2
+ \cdots
\label{Decoupling:ZA2}
\end{equation}
Finally, we arrive at
\begin{equation}
\zeta_A^{-1}(\mu) = 1 + \frac{4}{3} L \frac{\alpha(\mu)}{4\pi}
+ \left( - 4 L + \frac{13}{3} \right)
\left(\frac{\alpha(\mu)}{4\pi}\right)^2
+ \cdots
\label{Decoupling:zetaA2}
\end{equation}

Now we are going to compare the charge in QED
and in the low-energy theory (QPD).
In QED, due to Ward identities,
\begin{equation}
Z_\alpha = Z_A^{-1}\,,\qquad
Z_\alpha^{\text{os}} = \left(Z_A^{\text{os}}\right)^{-1}\,.
\label{Decoupling:Ward}
\end{equation}
The on-shell charge (measured at large distances)
is the same in both theories:
\begin{equation}
\alpha_{\text{os}} = \alpha'_{\text{os}}\,.
\label{Decoupling:erel}
\end{equation}
Therefore,
\begin{equation}
\alpha(\mu) = \zeta_\alpha(\mu) \alpha'(\mu)\,,
\label{Decoupling:erelMS}
\end{equation}
where
\begin{equation}
\zeta_\alpha(\mu) = \frac{Z_\alpha^{\text{os}}}{Z_\alpha}
= \zeta_A^{-1}(\mu)\,.
\label{Decoupling:zetae}
\end{equation}

At what scale $\mu$ should we do this matching?
In principle, this does not matter.
However, in practice, we know only a few terms in perturbative series.
Their coefficients contain powers of the logarithm $L$~(\ref{Decoupling:via}).
If it is large, truncating the series after some term
produces large errors.
Therefore, it is better to use $\mu\sim M$.
One popular choice is $\mu_0$ defined as the root of the equation
\begin{equation}
\mu_0 = M(\mu_0)
\label{Decoupling:mu0}
\end{equation}
(it corresponds to $L=0$).
Then~(\ref{Decoupling:zetaA2})
\begin{equation}
\zeta_\alpha(\mu_0) = 1 + \frac{13}{3}
\left(\frac{\alpha(\mu_0)}{4\pi}\right)^2 + \cdots
\label{Decoupling:zeta2a}
\end{equation}
Another popular choice is to use the on-shell mass:
$\mu=M_{\text{os}}$.
It is related to the \MS{} mass by
\begin{equation}
\frac{M(\mu)}{M_{\text{os}}} = 1
- 6 \left( \log\frac{\mu}{M_{\text{os}}} + \frac{2}{3} \right)
\frac{\alpha}{4\pi} + \cdots
\label{Decoupling:Mmu}
\end{equation}
and we obtain
\begin{equation}
\zeta_\alpha(M_{\text{os}}) = 1 + 15
\left(\frac{\alpha(M_{\text{os}})}{4\pi}\right)^2 + \cdots
\label{Decoupling:zeta2b}
\end{equation}
In general, for any $\mu$ which differs from the mass $M$
by $\mathcal{O}(\alpha)$, $\zeta_\alpha(\mu)$ differs from 1
by $\mathcal{O}(\alpha^2)$
(this is so, in particular, for $\mu_0$ and $M_{\text{os}}$).
If we choose, for example, $\mu=2M$ or $\mu=M/2$ instead,
we'll have $\zeta_\alpha(\mu)=1+\mathcal{O}(\alpha)$,
i.e., the correction will be much more important.
Therefore, it is better to match the full theory
and the low-energy one at some scale $\mu$
which differs from $M_{\text{os}}$ (or $\mu_0$)
by $\mathcal{O}(\alpha)$.

\section{QED with heavy muons}
\label{S:mu}

\subsection{Qedland}
\label{S:QED0}

Physicists in Qedland suspect that QED is also only
a low-energy effective theory.
We know that they are right, and muons exist%
\footnote{In our real world $M_\pi\sim M_\mu$;
for simplicity we shall assume that pions don't exist.}.
There are two ways in which they can search for new physics:
\begin{itemize}
\item by increasing the energy of their $e^+ e^-$ colliders
in the hope to produce pairs of new particles;
\item by performing high-precision experiments at low energies
(e.g., by measuring the electron magnetic moment).
\end{itemize}
New physics can produce new local interactions of photons,
electrons, and positrons,
which should be included in the effective QED Lagrangian%
\footnote{We were lucky that the scale of new physics in QED
is far away from the electron mass $m$.
Contributions of heavy-particle loops are also strongly suppressed
by powers of $\alpha$.
Therefore, the prediction for the electron magnetic moment
from the pure QED Lagrangian (without nonrenormalizable corrections)
is in good agreement with experiment.
After this spectacular success of the simplest Dirac equation
(without the Pauli term) for electrons,
physicists expected that the same holds for the proton,
and its magnetic moment is $e/(2 m_p)$.
No luck here.
This shows that the picture of the proton
as a point-like structureless particle
is a poor approximation at the energy scale $m_p$.}.

For simplicity, we shall first assume that electrons are massless.
This is a valid approximation if we want to describe phenomena
at energies much higher than $m_e$
but much lower than the new physics scale $M$
(in reality, the muon mass).
A non-zero electron mass will be re-instated in Sect.~\ref{S:mass}.
What operators of dimensions $>4$ can we add to the QED Lagrangian?
The first gauge-invariant operator appears at dimension 5
(the anomalous magnetic moment):
\begin{equation}
O = \bar{\psi} F_{\mu\nu} \sigma^{\mu\nu} \psi\,,
\label{QED0:mu}
\end{equation}
where $\sigma^{\mu\nu}=(i/2)[\gamma^\mu,\gamma^\nu]$.
However, it flips the electron helicity.
Helicity of a massless electron is conserved in QED.
If we assume that new charged particles interact with photons,
but not with electrons
(or, if they interact with electrons,
this interaction does not violate helicity conservation),
this operator cannot appear in the Lagrangian at $m_e=0$.

Under this assumption, new interactions first appear at dimension 6.
Electron--electron contact interactions
\begin{equation}
O_n = (\bar{\psi} \gamma_{(n)} \psi) (\bar{\psi} \gamma_{(n)} \psi)
\label{QED0:contact}
\end{equation}
(see~\ref{App:gamman}) conserve helicity at odd $n$.
The operators
\begin{equation*}
(\partial_{\mu} F^{\lambda\mu}) (\partial^\nu F_{\lambda\nu})\,,\qquad
\bar{\psi} \partial_\nu F^{\mu\nu} \gamma_\mu \psi
\end{equation*}
reduce to $O_1$~(\ref{QED0:contact}) due to equations of motion, and
\begin{equation*}
\bar{\psi} \partial_\lambda F_{\mu\nu} \gamma^{[\lambda} \gamma^\mu \gamma^{\nu]} \psi = 0\,.
\end{equation*}
Therefore, our first task is to investigate renormalization
of the operators~(\ref{QED0:contact}).

An important aspect of all effective field theories is power counting.
In this particular case it is trivial;
however, in more complicated situations, it becomes more involved.
When considering QED processes with small characteristic momenta $p$,
we have a small parameter $\lambda\sim p/M$.
The characteristic distance at which soft fields vary is large: $x\sim1/\lambda$,
and $\partial_\mu\sim\lambda$ when acting on soft fields.
The soft photon propagator is
\begin{equation*}
{<}0|T\left\{A_\mu(x) A_\nu(0)\right\}|0{>} \sim
\int \frac{d^4 p}{(2\pi)^4} e^{-i p\cdot x} \frac{1}{p^2}
\left[ g_{\mu\nu} - (1-a) \frac{p_\mu p_\nu}{p^2} \right]\,,
\end{equation*}
from $p\sim\lambda$ and $x\sim1/\lambda$ we obtain $A\sim\lambda$
(this means that $D_\mu \sim \lambda$ is homogeneous in $\lambda$).
The soft electron propagator is
\begin{equation*}
{<}0|T\left\{\psi(x) \bar{\psi}(0)\right\}|0{>} \sim
\int \frac{d^4 p}{(2\pi)^4} e^{-i p\cdot x} \frac{1}{\rlap/p}\,,
\end{equation*}
and we obtain $\psi\sim\lambda^{3/2}$.
The terms in the leading-order Lagrangian scale as
$F_{\mu\nu} F^{\mu\nu} \sim \lambda^4$,
$\bar{\psi} i\D \psi \sim \lambda^4$;
this means that the characteristic action is of order $1$,
as expected.
The first power corrections~(\ref{QED0:contact}) to the Lagrangian
scale as $\lambda^6$, and their contribution to the action is $\sim\lambda^2$.

Of course, we can add higher-dimensional contributions to the Lagrangian,
with further unknown coefficients.
To any finite order in $1/M$, the number of such coefficients is finite,
and the theory has predictive power.
For example, if we want to work at the order $1/M^4$,
then either a single $1/M^4$ (dimension 8) vertex
or two $1/M^2$ ones (dimension 6) can occur in a diagram.
UV divergences which appear in diagrams with two dimension 6 vertices
are compensated by dimension 8 counterterms.
So, the theory can be renormalized.
The usual arguments about nonrenormalizability
are based on considering diagrams with arbitrarily many vertices
of nonrenormalizable interactions (operators of dimensions $>4$);
this leads to infinitely many free parameters in the theory.
As stressed already, in any effective theory we always work
up to some finite order in $1/M$,
and the number of parameters is finite.

\subsection{Renormalization of four-fermion operators}
\label{S:4f}

First we'll consider a somewhat simpler problem~\cite{DG:91}:
renormalization of the operators
\begin{equation}
O^0_n = (\bar{\psi}_{10} \gamma_{(n)} \psi_{20})
(\bar{\psi}_{30} \gamma_{(n)} \psi_{40})
\label{4f:O0}
\end{equation}
with 4 different lepton flavours ($\gamma_{(n)}$ is defined in~(\ref{App:gamman})).
In 4 dimensions, only operators with $n\le4$ exist.
However, in dimensional regularization bare operators with all $n$ exist
(those with $n\ge4$ are called \emph{evanescent operators}).
If we use the standard minimal subtraction,
renormalized evanescent operators don't vanish.
Naturally, we want them to vanish,
and thus to have only a finite number of renormalized operators.
This can be achieved by tuning the renormalization constants
(which are no longer minimal in the sense of~(\ref{Decoupling:min})).

\begin{figure}[t]
\begin{center}
\begin{picture}(152,87)
\put(16,68){\makebox(0,0){\includegraphics{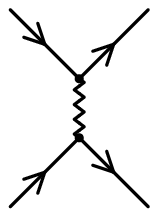}}}
\put(96,68){\makebox(0,0){\includegraphics{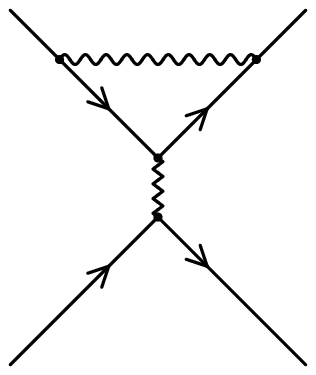}}}
\put(136,68){\makebox(0,0){\includegraphics{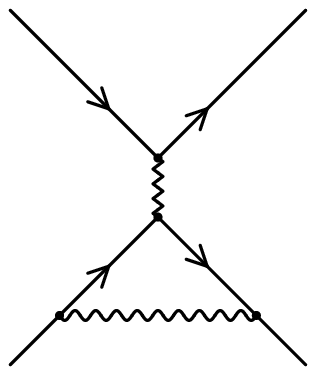}}}
\put(16,22){\makebox(0,0){\includegraphics{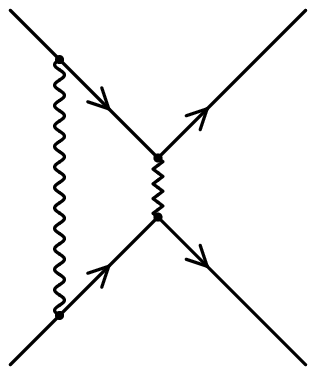}}}
\put(56,22){\makebox(0,0){\includegraphics{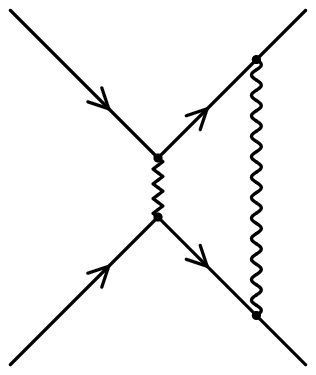}}}
\put(96,22){\makebox(0,0){\includegraphics{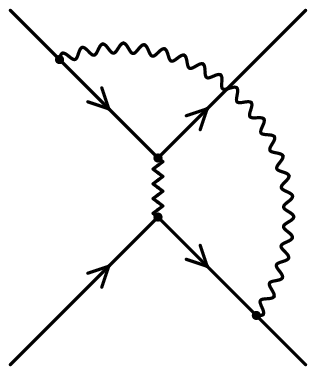}}}
\put(136,22){\makebox(0,0){\includegraphics{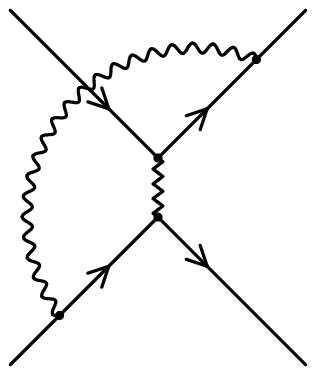}}}
\put(16,46){\makebox(0,0)[b]{$a$}}
\put(96,46){\makebox(0,0)[b]{$b$}}
\put(136,46){\makebox(0,0)[b]{$c$}}
\put(16,0){\makebox(0,0)[b]{$d$}}
\put(56,0){\makebox(0,0)[b]{$e$}}
\put(96,0){\makebox(0,0)[b]{$f$}}
\put(136,0){\makebox(0,0)[b]{$g$}}
\end{picture}
\end{center}
\caption{Matrix element of $O_n^0$}
\label{F:Eva}
\end{figure}

Let's calculate the matrix element of the bare operator~(\ref{4f:O0})
at one loop, keeping only the leading terms in the $\varepsilon$ expansion.
We'll write the tree contribution (Fig.~\ref{F:Eva}$a$%
\footnote{Here the zigzag line is actually a point;
it only shows which fermion legs are connected to the same $\gamma$ matrix.})
as $\gamma_{(n)}\otimes\gamma_{(n)}$;
it should be multiplied by the external legs renormalization factor
$\bigl(Z_\psi^{1/2}\bigr)^4$~(\ref{App:Zpsi}).
After averaging over $k$ directions and using~(\ref{App:UV}),
the contributions of Fig.~\ref{F:Eva}$b$, $c$; $d$, $e$; $f$, $g$ are
\begin{align*}
2&\left[ \frac{1}{d} \gamma_\mu \gamma_\nu \gamma_{(n)} \gamma^\nu \gamma^\mu \otimes \gamma_{(n)}
- (1-a) \gamma_{(n)} \otimes \gamma_{(n)} \right] \frac{\alpha}{4\pi\varepsilon}\\
{}-2&\left[ \frac{1}{d} \gamma_\mu \gamma_\nu \gamma_{(n)} \otimes \gamma^\mu \gamma^\nu \gamma_{(n)}
- (1-a) \gamma_{(n)} \otimes \gamma_{(n)} \right] \frac{\alpha}{4\pi\varepsilon}\\
{}+2&\left[ \frac{1}{d} \gamma_{(n)} \gamma_\nu \gamma_\mu \otimes \gamma^\mu \gamma^\nu \gamma_{(n)}
- (1-a) \gamma_{(n)} \otimes \gamma_{(n)} \right] \frac{\alpha}{4\pi\varepsilon}\,.
\end{align*}
The first of them is the same as for two-fermion currents,
and has the structure $\gamma_{(n)}\otimes\gamma_{(n)}$, see~(\ref{App:h}).

The other ones can be calculated using the identities
\begin{equation}
\begin{split}
&\gamma_\mu \gamma_{(n)} \otimes \gamma^\mu \gamma_{(n)} =
\gamma_{(n+1)} \otimes \gamma_{(n+1)} + n (d-n+1) \gamma_{(n-1)} \otimes \gamma_{(n-1)}\,,\\
&\gamma_{(n)} \gamma_\mu \otimes \gamma^\mu \gamma_{(n)} = (-1)^n \left[
\gamma_{(n+1)} \otimes \gamma_{(n+1)} - n (d-n+1) \gamma_{(n-1)} \otimes \gamma_{(n-1)} \right]\,.
\end{split}
\label{4f:h}
\end{equation}
Indeed,
\begin{equation*}
\gamma_{(n)} \otimes \gamma_{(n)} = n! \left( \gamma_0 \gamma_1 \cdots \gamma_{n-1} \otimes
\gamma^0 \gamma^1 \cdots \gamma^{n-1} + \cdots \right)
\end{equation*}
(there are $d!/(n!\,(d-n)!)$ terms in the sum).
Now let's multiply the left part (before $\otimes$) by $\gamma_\mu$
and the right one by $\gamma^\mu$, say, from the left.
There are two kinds of contributions.
When the value of the index $\mu$ is not one of those already present,
we get $n+1$ $\gamma$ matrices both before and after $\otimes$.
Each term is produced $n+1$ times; this factor converts $n!$ into $(n+1)!$,
and the sum of all such contributions is $\gamma_{(n+1)} \otimes \gamma_{(n+1)}$,
the first term in the right-hand side of the first identity.
Otherwise, one $\gamma$ matrix gets squared,
and we get $n-1$ $\gamma$ matrices both before and after $\otimes$.
There are $n$ such values of $\mu$ for each term;
the resulting bracket contains $d!/((n-1)!\,(d-n+1)!)$ terms,
so, each one is produced $d-n+1$ times.
Converting the common factor $n!$ to $(n-1)!$,
we get the second term in the first identity.
The second identity follows easily: contributions of the first kind
get an extra factor $(-1)^n$ from commuting $\gamma_\mu$ to the left;
contributions of the second kind get $(-1)^{n-1}$.
Note a very important property:
when $n=5$ (Dirac structure vanishing in 4 dimensions),
the contribution $\gamma_{(n-1)} \otimes \gamma_{(n-1)}$
(Dirac structure not vanishing in 4 dimensions)
comes with the factor $\sim\varepsilon$.

Combining all contributions, we obtain a gauge-invariant result
for the matrix element of the bare operator~(\ref{4f:O0}):
\begin{equation}
\begin{split}
{<}O_n^0{>} ={}& \left[1 + 2 (n-1) (n-3) \frac{\alpha}{4\pi\varepsilon}\right]
\gamma_{(n)} \otimes \gamma_{(n)}\\
&{}- \Bigl[ \gamma_{(n+2)} \otimes \gamma_{(n+2)}
+ n (n-1) (d-n+1) (d-n+2) \gamma_{(n-2)} \otimes \gamma_{(n-2)} \Bigr]
\frac{\alpha}{4\pi\varepsilon}\,.
\end{split}
\label{4f:M}
\end{equation}
In particular,
\begin{equation}
\left(
\begin{array}{c}
{<}O_1^0{>}\\
{<}O_3^0{>}\\
\hline
{<}O_5^0{>}\\
\vdots
\end{array}
\right)
= \left[ 1 +
\left(
\begin{array}{cc|cc}
0 & -1\\
-36 & 0 & -1\\
\hline
& 40 \varepsilon & 16 & -1\\
&& \vdots & \ddots
\end{array}
\right)
\frac{\alpha}{4\pi\varepsilon}
\right]
\left(
\begin{array}{c}
\gamma_{(1)}\otimes\gamma_{(1)}\\
\gamma_{(3)}\otimes\gamma_{(3)}\\
\hline
\gamma_{(5)}\otimes\gamma_{(5)}\\
\vdots
\end{array}
\right)
\label{4f:M1}
\end{equation}
(where the physical operators with $n\le4$ and the evanescent ones ($n\ge5$)
are separated by lines).
Here only the leading term in $\varepsilon$ is given
for each element of the matrix.
These leading terms originate from the UV divergent loop integral~(\ref{App:UV});
they don't depend on external momenta and masses.
The contribution $\gamma_{(3)}\otimes\gamma_{(3)}$ in the matrix element
of the bare evanescent operator $O_5^0$ acquires the factor $\varepsilon$
from~(\ref{4f:h}), and hence is finite, not $\mathcal{O}(1/\varepsilon)$.

The column vectors of the bare and renormalized operators are related
by the matrix of renormalization constants:
\begin{equation}
O^0 = Z(\alpha(\mu)) O(\mu)\,,\quad
O(\mu) = Z^{-1}(\alpha(\mu)) O^0\,.
\label{4f:Z}
\end{equation}
The renormalized operators obey the renormalization group equations
\begin{equation}
\frac{d O(\mu)}{d\log\mu} + \gamma(\alpha(\mu)) O(\mu) = 0\,,
\label{4f:RG}
\end{equation}
where the anomalous dimension matrix is
\begin{equation}
\gamma = Z^{-1} \frac{d Z}{d\log\mu} = - \frac{d Z^{-1}}{d\log\mu} Z\,.
\label{4f:gamma}
\end{equation}
We want the renormalized evanescent operators to vanish:
\begin{equation}
O(\mu) = \left(
\begin{array}{c}
O_1(\mu)\\
O_3(\mu)\\
\hline
0\\
\vdots
\end{array}
\right)\,.
\label{4f:reneva}
\end{equation}
Therefore, we have to include this $\mathcal{O}(1)$ term in $Z$:
\begin{equation}
Z(\alpha) = 1 +
\left(
\begin{array}{cc|cc}
0 & -1\\
-36 & 0 & -1\\
\hline
& 40 \varepsilon & 16 & -1\\
&& \vdots & \ddots
\end{array}
\right)
\frac{\alpha}{4\pi\varepsilon}\,.
\label{4f:Z1}
\end{equation}
Then
\begin{equation*}
O_5(\mu) = O_5^0 - 40 \frac{\alpha(\mu)}{4\pi} O_3^0 + \cdots
\end{equation*}
and the $\gamma_{(3)}\otimes\gamma_{(3)}$ contributions of $O_5^0$ and $O_3^0$
cancel in matrix elements of $O_5(\mu)=0$.
The renormalization constant matrix~(\ref{4f:Z1}) is not minimal:
it contains not only negative powers of $\varepsilon$,
but also $\mathcal{O}(1)$ contributions.
The one-loop anomalous dimension matrix is
\begin{equation}
\gamma(\alpha) = - 2
\left(
\begin{array}{cc|cc}
0 & -1\\
-36 & 0 & -1\\
\hline
&& 16 & -1\\
&& \vdots & \ddots
\end{array}
\right)
\frac{\alpha}{4\pi}\,.
\label{4f:gamma1}
\end{equation}
Its lower left block vanishes;
therefore, the form~(\ref{4f:reneva}) is preserved by evolution~(\ref{4f:RG}).
Evolution of the physical operators is determined by the upper left block.
When calculating this block at one loop,
one may forget about evanescent operators.

With the two-loop accuracy, the non-minimal renormalization matrix
can be written as
\begin{equation}
Z(\alpha) = 1
+ \left( Z_{10} + \frac{Z_{11}}{\varepsilon} \right) \frac{\alpha}{4\pi}
+ \left( Z_{20} + \frac{Z_{21}}{\varepsilon} + \frac{Z_{22}}{\varepsilon^2} \right)
\left(\frac{\alpha}{4\pi}\right)^2\,.
\label{4f:Zform}
\end{equation}
The anomalous dimension matrix~(\ref{4f:gamma}) must be finite at $\varepsilon\to0$.
This gives the self-consistency condition
\begin{equation}
Z_{22} = \frac{1}{2} Z_{11} (Z_{11} -\beta_0)\,,
\label{4f:self}
\end{equation}
i.\,e., the $1/\varepsilon^2$ part of the two-loop contribution $Z_{22}$
is determined by the $1/\varepsilon$ part of the one-loop contribution $Z_{11}$.
The anomalous dimension is
\begin{equation}
\gamma(\alpha) = - 2 Z_{11} \frac{\alpha}{4\pi}
- 2 (2 Z_{21} - Z_{10} Z_{11} - Z_{11} Z_{10} + \beta_0 Z_{10})
\left(\frac{\alpha}{4\pi}\right)^2\,.
\label{4f:gamma2}
\end{equation}
As we have already discussed, the one-loop $\mathcal{O}(\varepsilon^0)$ term
has the structure
\begin{equation}
Z_{10} = \left(
\begin{array}{cc}
0 & 0\\
a & 0
\end{array}
\right)\,;
\label{4f:z10}
\end{equation}
it originates from the product of $1/\varepsilon$ divergences of one-loop integrals
(which are momentum-independent) and the factor $\varepsilon$
from $\gamma$-matrix algebra.
Because of this factor, there are no $1/\varepsilon$ terms in the lower left corner:
\begin{equation}
Z_{11} = \left(
\begin{array}{cc}
b & c\\
0 & d
\end{array}
\right)\,.
\label{4f:z11}
\end{equation}
The lower left corner of
\begin{equation*}
Z_{21} = \left(
\begin{array}{cc}
e & f\\
g & h
\end{array}
\right)
\end{equation*}
comes from the product of $1/\varepsilon^2$ divergences of two-loop integrals
(which are momentum-independent) and the factor $\varepsilon$
from $\gamma$-matrix algebra.
These $1/\varepsilon^2$ divergences of two-loop integrals are determined
by products of $1/\varepsilon$ divergences of one-loop integrals.
Namely, the lower left corner of the anomalous dimension matrix~(\ref{4f:gamma2})
must vanish, in order to preserve the form~(\ref{4f:reneva}).
This leads to the self-consistency condition
\begin{equation}
g = \frac{1}{2} (ab + da - \beta_0 a)\,.
\label{4f:selfeva}
\end{equation}
Evolution~(\ref{4f:RG}) of the physical (non-evanescent) operators is determined
by the upper left corner of the anomalous dimension matrix $\gamma$~(\ref{4f:gamma2});
the two-loop contribution to it is
\begin{equation*}
- 2 (2 e + ca) \left(\frac{\alpha}{4\pi}\right)^2\,.
\end{equation*}
In order to find it, one needs $e$, the $1/\varepsilon$ part of two-loop diagrams
with the insertion of a physical operator.
Such a calculation (though in a more complicated case)
is explained in~\cite{CMM:97}.
The best way is to nullify all external momenta,
and to insert masses into all denominators to regularize IR divergences.

\begin{figure}[t]
\begin{center}
\begin{picture}(54,44)
\put(11,22){\makebox(0,0){\includegraphics{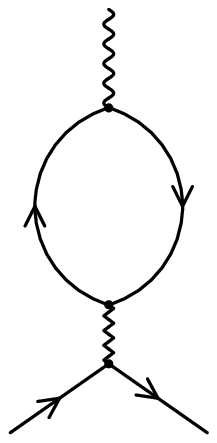}}}
\put(43,22){\makebox(0,0){\includegraphics{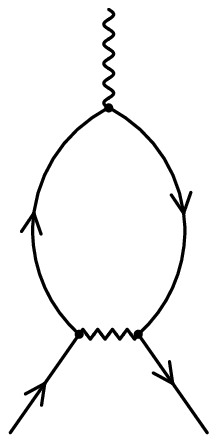}}}
\end{picture}
\end{center}
\caption{Matrix element of $O_n^0$}
\label{F:Eva2}
\end{figure}

In the case of the operators~(\ref{QED0:contact}) with identical flavours,
an electron line leaving the operator vertex can returns to the same vertex.
The $e^+e^-\gamma$ matrix element at one loop
is given by the diagrams in Fig.~\ref{F:Eva2}.
These diagrams contain the electron loop
\begin{equation}
\int \frac{d^d k}{(2\pi)^d}\,
i \frac{\rlap/k+\rlap/p}{(k+p)^2}\,
i e_0 \gamma^\mu\,
i \frac{\rlap/k}{k^2}
= - i e_0 \left( p^2 \gamma^\mu - \rlap/p p^\mu \right)
\frac{1}{2}\,\frac{d-2}{d-1}
\int \frac{d^d k}{(2\pi)^d}\,
\frac{1}{k^2 (k+p)^2}
\label{4f:floop}
\end{equation}
which vanishes after contraction with $p_\mu$.
Therefore, this matrix element is proportional to the tree-level matrix element
of the operator $O=\bar{\psi}\partial_\nu F^{\mu\nu}\gamma_\mu\psi$.
The first diagram contributes only to ${<}O_1^0{>}$.
Taking the trace of~(\ref{4f:floop}) with a single $\gamma$ matrix
(and the fermion loop factor $-1$)
we recover the familiar expression for the one-loop photon self energy.
The divergent part of the matrix element is
\begin{equation}
{<}O_1^0{>}_1 = - \frac{8}{3} \frac{e}{(4\pi)^2\varepsilon} {<}O{>}
\label{4f:O1}
\end{equation}
(we have taken into account that there are 2 such contributions).

The second diagram for ${<}O_n^0{>}$ contains
\begin{equation}
\gamma_{(n)} \gamma^\mu \gamma_{(n)} = A_n \gamma^\mu\,.
\label{4f:Andef}
\end{equation}
Multiplying this by $\gamma_\mu$ and using~(\ref{App:h}) we have
\begin{equation*}
A_n = (-1)^n \frac{d-2n}{d} \gamma_{(n)} \gamma_{(n)}\,.
\end{equation*}
There are $d!/(n!(d-n)!)$ ways to select $n$ indices of $d$,
and they contribute equally to $\gamma_{(n)} \gamma_{(n)}$:
\begin{equation*}
\gamma_{(n)} \gamma_{(n)} = \frac{d!}{n!(d-n)!} \gamma_0 \cdots \gamma_{n-1}
\left( \gamma^0 \cdots \gamma^{n-1} \pm \cdots \right)\,.
\end{equation*}
There are $n!$ terms in the bracket; they contribute equally, namely,
$(-1)^{n(n-1)/2}$ each.
Finally, we obtain
\begin{equation}
\begin{split}
A_n &= (-1)^{n(n+1)/2} (d-2n) \frac{\Gamma(d)}{\Gamma(d-n+1)}\\
&= (-1)^{n(n+1)/2} (d-2n) \cdot (d-1) (d-2) \cdots (d-n+1)
\end{split}
\label{4f:An}
\end{equation}
(naturally, $A_n$ contains $(d-4)$ when $n\ge5$).
Using this algebraic fact, we arrive at the divergent part of the two contributions
given by the second diagram in Fig.~\ref{F:Eva2}:
\begin{equation}
{<}O_n^0{>}_2 = - (-1)^{n(n+1)/2} 8 \frac{n-2}{(4-n)!}
\frac{e}{(4\pi)^2\varepsilon} {<}O{>}\,.
\label{4f:O2}
\end{equation}

Due to the equation of motion, $O=eO_1$.
Collecting all contributions together, we obtain
\begin{equation*}
\left(
\begin{array}{c}
{<}O_1^0{>}\\
{<}O_3^0{>}
\end{array}
\right)
= \left[ 1 +
\left(
\begin{array}{cc}
-4 & -1\\
-37 & 0
\end{array}
\right) \frac{\alpha}{4\pi\varepsilon} \right]
\left(
\begin{array}{c}
{<}O_1^0{>}_{\text{tree}}\\
{<}O_3^0{>}_{\text{tree}}
\end{array}
\right)\,,
\end{equation*}
where only $1/\varepsilon$ parts of the one-loop matrix elements are retained.
This gives the anomalous dimension matrix
\begin{equation}
\gamma = 2
\left(
\begin{array}{cc}
4 & 1\\
37 & 0
\end{array}
\right) \frac{\alpha}{4\pi}
\label{4f:gamma0}
\end{equation}
of the operators~(\ref{QED0:contact}).

\subsection{Contact interaction of electrons}
\label{S:Cont}

The Lagrangian of the effective low-energy QED (without muons)
with the $1/M^2$ accuracy is
\begin{equation}
L = L_0 + L_1\,,\quad
L_1 = c_1^0 O_1^0 + c_3^0 O_3^0 = c_1(\mu) O_1(\mu) + c_3(\mu) O_3(\mu)\,,
\label{Cont:L}
\end{equation}
where $L_0$ is the standard massless QED Lagrangian,
the bare operators $O_n^0$ are defined by
\begin{equation}
O_n^0 = (\bar{\psi}_0 \gamma_{(n)} \psi_0) (\bar{\psi}_0 \gamma_{(n)} \psi_0)\,,
\label{Cont:O0}
\end{equation}
and the coefficients $c_i^0\sim1/M^2$ play the role of charges
at the vertices produced by these operators
(similarly to $e_0$ at the ordinary electron--photon vertices).

Interaction terms (of some dimensionality) in any effective Lagrangian can be written as
\begin{equation}
L_1 = c_0^T O_0 = c^T(\mu) O(\mu)\,,\quad
O_0 = Z(\alpha(\mu)) O(\mu)\,,\quad
c(\mu) = Z^T(\alpha(\mu)) c_0\,,
\label{Cont:L1}
\end{equation}
where $O_0$ is the column vector of the bare operators,
$c_0^T$ is the row vector of the bare couplings ($T$ means transposition),
and $Z$ is the matrix of renormalization constants of these operators.
The renormalized couplings (called also Wilson coefficients)
obey the renormalization group equations
\begin{equation}
\frac{d c(\mu)}{d\log\mu} = \gamma^T(\alpha(\mu)) c(\mu)\,,
\label{Cont:RG}
\end{equation}
where the anomalous dimension matrix of the operators $O$
is defined by~(\ref{4f:gamma}).
The Wilson coefficients $c_i(\mu_0)$
are determined by matching --- equating some $S$-matrix elements in the full theory
(expanded in $p_i/M$) and in the effective theory.
It is most convenient to use some $\mu_0\sim M$;
then $c_i(\mu_0)$ are given by perturbative series in $\alpha(\mu_0)$
containing no large logarithms.
They contain all the information about physics at the scale $M$
which is important for low-energy processes.
The Wilson coefficients $c_i(\mu)$ at low normalization scales $\mu$
are obtained by solving the RG equations~(\ref{Cont:RG})
with the initial conditions at $\mu_0\sim M$ given by matching.
The effective theory knows nothing about $M$;
the only information about it is contained in $c_i(\mu)$.
When the effective Lagrangian is applied to some physical process
with small momenta $p_i\ll M$,
it is most convenient to use $\mu$ of the order of the characteristic momenta:
then the results will contain no large logarithms.
This solution of the RG equation sums large logarithmic terms
in perturbation series.

As already discussed, the anomalous dimension matrix has the block structure
\begin{equation*}
\gamma = \left(\begin{array}{cc}*&*\\0&*\end{array}\right) \Rightarrow
\gamma^T = \left(\begin{array}{cc}*&0\\{}*&*\end{array}\right)
\end{equation*}
in terms of physical and evanescent operators.
This means that the evolution equations for Wilson coefficients of physical operators
don't involve those for evanescent operators.
The later Wilson coefficients are non-zero;
but this does not matter, because renormalized evanescent operators vanish.

\begin{figure}[b]
\begin{center}
\begin{picture}(26,22)
\put(13,11){\makebox(0,0){\includegraphics{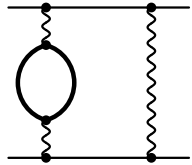}}}
\end{picture}
\end{center}
\caption{A diagram having the $\gamma_{(3)}\otimes\gamma_{(3)}$ structure}
\label{F:O3}
\end{figure}

As we have seen in Sect.~\ref{S:Coulomb},
the $q^2$ term in the muon loop~(\ref{Coulomb:Pi1}) leads to the contribution
\begin{equation}
\Delta L = c O\,,\quad
c = - \frac{2}{15} \frac{\alpha}{4\pi} \frac{1}{M^2}
+ \mathcal{O}(\alpha^2)\,,\quad
O = (\partial^\mu F_{\lambda\mu}) (\partial_\nu F^{\lambda\nu})
\label{Cont:DL}
\end{equation}
to the Lagrangian.\
Due to the equation of motion,
\begin{equation}
O = e^2 O_1\,,
\label{Cont:EOM}
\end{equation}
where $O_1$ is defined by~(\ref{QED0:contact}).
Therefore,
\begin{equation}
c_1(M) = - \frac{2}{15} \frac{\alpha^2(M) + \mathcal{O}(\alpha^3)}{M^2}\,.
\label{Cont:c1}
\end{equation}
We can also understand this more directly
by matching the on-shell electron--electron scattering amplitudes:
\begin{equation}
\begin{split}
\raisebox{-9.3mm}{\includegraphics{eva.eps}} &=
\raisebox{-17mm}{\includegraphics{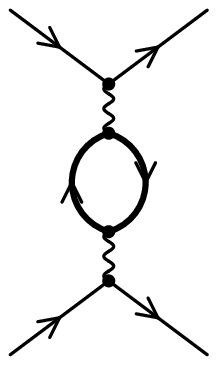}}\,,\\
2 i c_1^0 &= i \frac{e_0^2}{q^2}
\frac{4}{3} \frac{e_0^2 M_0^{-2\varepsilon}}{(4\pi)^{d/2}}
\Gamma(\varepsilon) \frac{d-4}{10} \frac{q^2}{M_0^2}
\Rightarrow - \frac{4}{15} i \frac{e_0^4}{(4\pi)^{d/2}} \frac{1}{M_0^{2+2\varepsilon}}\,.
\end{split}
\label{Cont:Match1}
\end{equation}
In order to obtain the structure $\gamma_{(3)}\otimes\gamma_{(3)}$,
we should have at least three $\gamma$ matrices along each electron line.
Such diagrams first appear at two loops (Fig.~\ref{F:O3}).
Therefore,
\begin{equation}
c_3(M) = \frac{\mathcal{O}(\alpha^3(M))}{M^2}\,.
\label{Cont:c3}
\end{equation}

It is convenient to rewrite the RG equation~(\ref{Cont:RG}) in the form
\begin{equation}
\frac{d c}{d\log\alpha} = - \frac{\gamma^T(\alpha)}{2\beta(\alpha)} c\,.
\label{Cont:RGalpha}
\end{equation}
At the leading (one-loop) order
\begin{equation*}
\beta(\alpha) = \beta_0 \frac{\alpha}{4\pi}\,,\quad
\gamma^T(\alpha) = \gamma^T_0 \frac{\alpha}{4\pi}\,.
\end{equation*}
The solution of this leading-order equation can be written as a matrix exponent:
\begin{equation}
c(\mu) = \left(\frac{\alpha(\mu)}{\alpha(M)}\right)^{-\gamma^T_0/(2\beta_0)} c(M)\,.
\label{Cont:mexp}
\end{equation}
If eigenvectors $v_i$ of $\gamma^T_0$,
\begin{equation*}
\gamma^T_0 v_i = \lambda_i v_i\,,
\end{equation*}
form a full basis%
\footnote{In some rare cases, there are not enough eigenvectors --- Jordan blocks of sizes $\ge2$ appear.
It is not difficult to solve RG equations in such cases, too.},
the solution~(\ref{Cont:mexp}) has the form
\begin{equation}
c(\mu) = \sum A_i \left(\frac{\alpha(\mu)}{\alpha(M)}\right)^{-\lambda_i/(2\beta_0)} v_i\,,
\label{Cont:eigen}
\end{equation}
where the constants $A_i$ are determined by the initial conditions:
\begin{equation}
c(M) = \sum A_i v_i\,.
\label{Cont:init}
\end{equation}
In our particular case
\begin{equation*}
\gamma^T_0 = 2
\left(
\begin{array}{cc}
4&37\\
1&0
\end{array}
\right)\,,\quad
\lambda_\pm = 2 \left( 2 \pm \sqrt{41} \right)\,,\quad
v_\pm =
\left(
\begin{array}{c}
2\pm\sqrt{41}\\
1
\end{array}
\right)\,,
\end{equation*}
and the solution of the leading-order RG equation
with the initial condition~(\ref{Cont:c1}) is
\begin{equation}
\begin{split}
\left(
\begin{array}{c}
c_1(\mu)\\
c_3(\mu)
\end{array}
\right)
= - \frac{1}{15\sqrt{41}} \frac{\alpha^2(M)}{M^2}
\Biggl[&
\left(\frac{\alpha(\mu)}{\alpha(M)}\right)^{\frac{3}{4}(2+\sqrt{41})}
\left(\begin{array}{c}2+\sqrt{41}\\1\end{array}\right)\\
&{} - \left(\frac{\alpha(\mu)}{\alpha(M)}\right)^{\frac{3}{4}(2-\sqrt{41})}
\left(\begin{array}{c}2-\sqrt{41}\\1\end{array}\right)
\Biggr]
\end{split}
\label{Cont:LL}
\end{equation}
(recall $\beta_0=-4/3$).

\subsection{Decoupling: photon field}
\label{S:photon}

Now we shall discuss the relation between the full theory
and the effective one more systematically.
Our full theory is QED with massless electrons
and muons having mass $M$.
When we consider processes with characteristic energies $E\ll M$,
the existence of muons is not important.
Everything can be described by an effective low-energy theory,
in which there are no muons.
In other words, muons only exist in loops of size $\sim1/M$;
if we are interested in processes having characteristic distances
much larger than $1/M$, such loops can be replaced by
local interactions of electrons and photons.

The effective low-energy theory contains the light fields ---
electrons and photons.
The Lagrangian of this theory, describing interactions of these fields
at low energies, contains all operators constructed from these fields
which are allowed by the symmetries.
Operators with dimensionalities $d_i>4$ are multiplied by coefficients
having negative dimensionalities;
these coefficients contain $1/M^{d_i-4}$.
Therefore, this Lagrangian can be viewed as an expansion in $1/M$.
The coefficients in this Lagrangian are fixed by matching ---
equating $S$-matrix elements up to some power of $p_i/M$.

Operators of the full theory are also expansions in $1/M$,
in terms of all operators of the effective theory
with appropriate quantum numbers.
In particular, the bare electron and the photon fields of the full theory are,
up to $1/M^2$ corrections,
\begin{equation}
\psi_0 = \left(\zeta_\psi^0\right)^{1/2} \psi'_0\,,\quad
A_0 = \left(\zeta_A^0\right)^{1/2} A'_0
\label{photon:fields}
\end{equation}
(primed quantities are those in the effective theory).
The bare parameters in the Lagrangians of the two theories are related by
\begin{equation}
e_0 = \left(\zeta_\alpha^0\right)^{1/2} e'_0\,,\quad
a_0 = \zeta_A^0 a'_0\,.
\label{photon:params}
\end{equation}
The \MS{} renormalized fields and parameters are related by
\begin{equation}
\begin{split}
&\psi(\mu) = \zeta_\psi^{1/2}(\mu) \psi'(\mu)\,,\quad
A(\mu) = \zeta_A^{1/2}(\mu) A'(\mu)\,,\\
&\alpha(\mu) = \zeta_\alpha(\mu) \alpha'(\mu)\,,\quad
a(\mu) = \zeta_A(\mu) a'(\mu)\,,
\end{split}
\label{photon:renorm}
\end{equation}
where
\begin{equation}
\zeta_\psi(\mu) =
\frac{Z'_\psi(\alpha'(\mu),a'(\mu))}{Z_\psi(\alpha(\mu),a(\mu))}
\zeta_\psi^0\,,\quad
\zeta_A(\mu) = \frac{Z'_A(\alpha'(\mu))}{Z_A(\alpha(\mu))} \zeta_A^0\,,\quad
\zeta_\alpha(\mu) =
\frac{Z'_\alpha(\alpha'(\mu))}{Z_\alpha(\alpha(\mu))} \zeta_\alpha^0\,.
\label{photon:renorm2}
\end{equation}

The photon propagators in the two theories are related by
\begin{equation}
\begin{split}
&D_\bot(p^2) \left(g_{\mu\nu} - \frac{p_\mu p_\nu}{p^2}\right)
+ a_0 \frac{p_\mu p_\nu}{(p^2)^2}\\
&{}= \zeta_A^0 \left[ D'_\bot(p^2) \left(g_{\mu\nu} - \frac{p_\mu p_\nu}{p^2}\right)
+ a'_0 \frac{p_\mu p_\nu}{(p^2)^2} \right]
+ \mathcal{O}\left(\frac{1}{M^2}\right)\,.
\end{split}
\label{photon:D}
\end{equation}
This explains why the same decoupling constant $\zeta_A$
describes decoupling for both the photon field $A$
and the gauge-fixing parameter $a$.
It is most convenient to do matching at $p^2\to0$,
then the power-suppressed terms in~(\ref{photon:D}) play no role.
The full-theory propagator near the mass shell is
\begin{equation}
D_\bot(p^2) = \frac{Z_A^{\text{os}}}{p^2}\,,\quad
Z_A^{\text{os}} = \frac{1}{1-\Pi(0)}\,.
\label{photon:full}
\end{equation}
Only diagrams with muon loops contribute to $\Pi(0)$,
all the other diagrams contain no scale.
In the effective theory
\begin{equation}
D'_\bot(p^2) = \frac{Z_A^{\prime\text{os}}}{p^2}\,,\quad
Z_A^{\prime\text{os}} = \frac{1}{1-\Pi'(0)} = 1\,,
\label{photon:eff}
\end{equation}
because all diagrams for $\Pi'(0)$ vanish.
Therefore,
\begin{equation}
\zeta_A^0 = \frac{Z_A^{\text{os}}}{Z_A^{\prime\text{os}}} = \frac{1}{1-\Pi(0)}\,.
\label{photon:zeta0}
\end{equation}

The calculation is essentially the same as in Sect.~\ref{S:Decoupling}.
At one loop the bare decoupling coefficient is given by~(\ref{Decoupling:zetaA01}).
For generality, let's suppose there are $n_l$ light lepton ``flavours''
in the effective theory and $n_f=n_l+1$ ``flavours'' in the full one.
Then
\begin{equation}
Z_A = 1 - \frac{4}{3} n_f \frac{\alpha(\mu)}{4\pi\varepsilon} + \cdots\,,\qquad
Z'_A = 1 - \frac{4}{3} n_l \frac{\alpha'(\mu)}{4\pi\varepsilon} + \cdots\,,
\label{photon:ZA1}
\end{equation}
because $\Pi(p^2)$ is proportional to the number of flavours.
With this accuracy, we may put $\alpha'(\mu)=\alpha(\mu)$,
and the renormalized decoupling coefficient is given
by~(\ref{Decoupling:zetaA1}).
It does not depend on $n_l$
(the case $n_l=0$ has been considered in Sect.~\ref{S:Decoupling}).

In order to calculate $\zeta_A$~(\ref{photon:renorm2}) with two-loop accuracy
we need $Z_A$ and $Z_A'$.
What is the \MS{} renormalization constant $Z_A$ with $n_f$ flavours?
It does not depend on masses.
If we suppose that all flavours have the same mass $M$,
then $\Pi(0)$ is given by the formula similar to~(\ref{Decoupling:Pi2})
where both the one-loop term and the two-loop one are multiplied by $n_f$
(starting from 3 loops, terms with different powers of $n_f$ appear).
Re-expressing via renormalized quantities at $\mu=\mu_0$~(\ref{Decoupling:mu0})
(with $Z_\alpha=Z_A^{-1}$, see~(\ref{photon:ZA1}), and $Z_m$~(\ref{Decoupling:Zm1}))
we have
\begin{equation*}
\left(Z_A^{\text{os}}\right)^{-1} = 1 - \Pi(0) = 1
+ \frac{4}{3} n_f \frac{\alpha(\mu_0)}{4\pi\varepsilon}
+ \left( \frac{16}{9} n_f + 2 \varepsilon \right)
\left(\frac{\alpha(\mu_0)}{4\pi\varepsilon}\right)^2
\end{equation*}
(unneeded higher powers of $\varepsilon$ are omitted).
The inverse quantity $Z_A^{\text{os}}$
should be equal to the minimal renormalization constant $Z_A(\alpha(\mu_0))$
times an expression finite at $\varepsilon\to0$, and we obtain
\begin{equation}
Z_A(\alpha) = 1 - \frac{4}{3} n_f \frac{\alpha}{4\pi\varepsilon}
- 2 \varepsilon n_f \left(\frac{\alpha}{4\pi\varepsilon}\right)^2\,.
\label{photon:ZA}
\end{equation}
The renormalization constant $Z_A'$ in the effective theory
contains $n_l$ instead of $n_f$ and $\alpha'$ instead of $\alpha$,
where $\alpha(\mu)$ and $\alpha'(\mu)$ are related by~(\ref{photon:renorm}),
and $\zeta_\alpha(\mu)=\zeta_A^{-1}(\mu)$
(this statement will be established in Sect.~\ref{S:charge};
at the moment, we only need it at one loop,
see~(\ref{Decoupling:zetaA1})).

With two-loop accuracy, the bare decoupling coefficient~(\ref{photon:zeta0})
is given by~(\ref{Decoupling:zA0}), because only diagrams with a muon loop
contribute to $\Pi(0)$.
The renormalized decoupling coefficient~(\ref{photon:renorm2})
is given by~(\ref{Decoupling:zetaA2}); it does not depend on $n_l$
(the case $n_l=0$ has been considered in Sect.~\ref{S:Decoupling}).
Starting from three loops, $n_l$-dependent terms appear.

The renormalization group equation
\begin{equation}
\frac{d\log\zeta_A(\mu)}{d\log\mu}
+ \gamma_A(\alpha(\mu)) - \gamma'_A(\alpha'(\mu)) = 0
\label{photon:RG}
\end{equation}
can be used to find $\mu$-dependence of the matching coefficient.
As discussed in Sect.~\ref{S:Decoupling},
it is better to impose the initial condition at $\mu=M$,
or at any $\mu$ which differs from $M$ by corrections of order $\alpha$.
Then $\zeta_A$ at this initial $\mu$ is $1+\mathcal{O}(\alpha^2)$,
i.e.\ corrections are small.
One popular choice is $\mu_0$~(\ref{Decoupling:mu0}),
with $\zeta_A(\mu_0)$ given by~(\ref{Decoupling:zeta2a}).

\subsection{Decoupling: electron field}
\label{S:electron}

\begin{figure}[ht]
\begin{center}
\begin{picture}(52,20)
\put(26,10){\makebox(0,0){\includegraphics{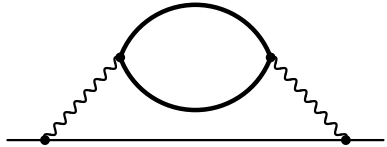}}}
\end{picture}
\end{center}
\caption{Two-loop diagram for $\Sigma_V(0)$}
\label{F:electron}
\end{figure}

The electron propagators in the full theory
and in the low-energy theory are related by
\begin{equation}
\rlap/p S(p) = \zeta_\psi^0\,\rlap/p S'(p)
+ \mathcal{O}\left(\frac{p^2}{M^2}\right)\,.
\label{electron:S}
\end{equation}
It is most convenient to do matching at $p\to0$,
where power corrections play no role.
The full-theory propagator near the mass shell is
\begin{equation}
S(p) = \frac{Z_\psi^{\text{os}}}{\rlap/p}\,,\quad
Z_\psi^{\text{os}} = \frac{1}{1-\Sigma_V(0)}\,,
\label{electron:full}
\end{equation}
where the self-energy of the massless electron is $\Sigma(p)=\Sigma_V(p^2)\rlap/p$.
Only diagrams with muon loops contribute to $\Sigma_V(0)$,
all the other diagrams contain no scale;
such diagrams first appear at two loops (Fig.~\ref{F:electron}).
In the effective theory
\begin{equation}
S'(p) = \frac{Z_\psi^{\prime\text{os}}}{\rlap/p}\,,\quad
Z_\psi^{\prime\text{os}} = \frac{1}{1-\Sigma_V'(0)} = 1\,,
\label{electron:eff}
\end{equation}
because all diagrams for $\Sigma_V'(0)$ vanish.
Therefore,
\begin{equation}
\zeta_\psi^0 = \frac{Z_\psi^{\text{os}}}{Z_\psi^{\prime\text{os}}}
= \frac{1}{1-\Sigma_V(0)}\,.
\label{electron:zeta0}
\end{equation}

At two loops (Fig.~\ref{F:electron}),
\begin{equation}
- i \rlap/p \Sigma_V(p^2) = \int \frac{d^d k}{(2\pi)^d}
i e_0 \gamma^\mu i \frac{\rlap/k+\rlap/p}{(k+p)^2} i e_0 \gamma^\nu
\left(\frac{-i}{k^2}\right)^2
i (k^2 g_{\mu\nu} - k_\mu k_\nu) \Pi(k^2)\,,
\label{electron:Sigma}
\end{equation}
where $i (k^2 g_{\mu\nu} - k_\mu k_\nu) \Pi(k^2)$
is the muon-loop contribution to the photon self-energy~(\ref{Coulomb:Pi}).
It is transverse; therefore, longitudinal parts of the photon propagators
($\sim k_\alpha k_\beta$) do not contribute,
and the result is gauge invariant.
We only need the linear term in $p$ in both sides:
\begin{equation*}
\begin{split}
\rlap/p \Sigma_V(0) &{}= - i e_0^2 \int \frac{d^d k}{(2\pi)^d}
\gamma^\mu (k^2 \rlap/p - 2 p\cdot k\,\rlap/k) \gamma^\nu
(k^2 g_{\mu\nu} - k_\mu k_\nu) \frac{\Pi(k^2)}{(k^2)^4}\\
&{}= - i e_0^2 \int \frac{d^d k}{(2\pi)^d}
\left[ \gamma_\mu (k^2 \rlap/p - 2 p\cdot k\,\rlap/k) \gamma^\mu
- \rlap/k \rlap/p \rlap/k + 2 p\cdot k\,\rlap/k \right]
\frac{\Pi(k^2)}{(k^2)^3}\,.
\end{split}
\end{equation*}
Averaging over $k$ directions by $p\cdot k\,\rlap/k\Rightarrow(k^2/d)\rlap/p$,
we obtain
\begin{equation}
\Sigma_V(0) = - i e_0^2 \frac{(d-1)(d-4)}{d}
\int \frac{d^d k}{(2\pi)^d} \frac{\Pi(k^2)}{(-k^2)^2}\,.
\label{electron:Sigma1}
\end{equation}

Let's calculate the integral
\begin{equation}
\int \frac{d^d k}{(2\pi)^d} \frac{\Pi(k^2)}{(-k^2)^n}
= i \frac{e_0^2 M_0^{4-2n-4\varepsilon}}{(4\pi)^d} I_n\,.
\label{electron:Idef}
\end{equation}
Using~(\ref{Coulomb:Pi1i}) (and setting $M_0=1$)
we can reduce it to the vacuum integrals~(\ref{Qedland:V2}) (Fig.~\ref{F:V2}):
\begin{equation}
I_n = \frac{2}{d-1} \left[4 V(1,1,n+1) - (d-2) V(1,1,n)\right]\,.
\label{electron:In}
\end{equation}
In particular,
\begin{equation}
I_2 = -  \Gamma^2(\varepsilon) \frac{2(d-6)}{(d-2)(d-5)(d-7)}\,.
\label{electron:I}
\end{equation}

Therefore, we obtain from~(\ref{electron:zeta0})
\begin{equation}
\zeta_\psi^0 = Z_\psi^{\text{os}} = 1 + \frac{e_0^4 M_0^{-4\varepsilon}}{(4\pi)^d} \Gamma^2(\varepsilon)
\frac{2(d-1)(d-4)(d-6)}{d(d-2)(d-5)(d-7)}\,.
\label{electron:zeta0res}
\end{equation}
The renormalized decoupling coefficient is
\begin{equation}
\zeta_\psi(\mu) = \zeta_\psi^0
\frac{Z_\psi'(\alpha'(\mu),a'(\mu))}{Z_\psi(\alpha(\mu),a(\mu))}\,.
\label{electron:zeraren}
\end{equation}
Its $\mu$-dependence can always be found
by solving the RG equation.
It is sufficient to obtain it at one point,
at some specific $\mu\sim M$,
to have the initial condition.
The most convenient point is $\mu=M$,
because $\alpha(M)=\alpha'(M)+\mathcal{O}(\alpha^3)$ (Sect.~\ref{S:charge})
and $a(M)=a'(M)+\mathcal{O}(\alpha^2)$ (Sect.~\ref{S:photon}),
and the differences can be neglected with our accuracy.
The renormalization constant $Z_\psi$ up to two loops has the form
\begin{equation}
Z_\psi(\alpha,a) = 1
- \frac{1}{2} \gamma_{\psi0} \frac{\alpha}{4\pi\varepsilon}
+ \frac{1}{8} \left[ \gamma_{\psi0} (\gamma_{\psi0}+2\beta_0)
  + \gamma_{A0} \gamma_{\psi0}'' a
  - 2 \gamma_{\psi1} \varepsilon \right]
\left(\frac{\alpha}{4\pi\varepsilon}\right)^2
+ \cdots
\label{electron:Zpsi}
\end{equation}
where
\begin{equation*}
\gamma_\psi = \frac{d\,\log Z_\psi}{d\,\log\mu}
= \gamma_{\psi0} \frac{\alpha}{4\pi\varepsilon}
+ \gamma_{\psi1} \left(\frac{\alpha}{4\pi\varepsilon})\right)^2
+ \cdots
\end{equation*}
is the anomalous dimension of the electron field,
$\gamma_{\psi0}=\gamma_{\psi0}'+\gamma_{\psi0}''a$,
and $\gamma_A$ is the anomalous dimension of the photon field.
In QED with $n_f$ lepton flavours
\begin{equation}
\gamma_\psi(\alpha,a) = 2 a \frac{\alpha}{4\pi}
- (4 n_f + 3) \left(\frac{\alpha}{4\pi}\right)^2 + \cdots
\label{electron:gamma}
\end{equation}
The effective theory renormalization constant $Z_\psi'$
is given by a similar formula with primed coefficients.
Their ratio is
\begin{equation}
\frac{Z_\psi(\alpha,a)}{Z_\psi'(\alpha',a')} = 1 + \frac{1}{4}
\left( \gamma_{\psi0} \Delta\beta_0
+ \frac{1}{2} \Delta\gamma_{A0} \gamma_{\psi0}'' a
- \Delta\gamma_{\psi1} \varepsilon \right)
\left(\frac{\alpha}{4\pi\varepsilon}\right)^2\,,
\label{electron:Zratio}
\end{equation}
where
\begin{equation*}
\Delta\beta_0 = -\frac{4}{3}\,,\quad
\Delta\gamma_{A0} = \frac{8}{3}\,,\quad
\Delta\gamma_{\psi1} = -4
\end{equation*}
are the single-flavour contributions to $\beta_0$, $\gamma_{A0}$, $\gamma_{\psi1}$.
We obtain
\begin{equation*}
\frac{Z_\psi(\alpha,a)}{Z_\psi'(\alpha,a)} = 1
+ \varepsilon \left(\frac{\alpha}{4\pi\varepsilon}\right)^2\,.
\end{equation*}
Re-expressing~(\ref{electron:zeta0res}) via the renormalized $\alpha(M)$,
\begin{equation*}
\zeta_\psi^0 = 1 + \varepsilon \left(1 - \frac{5}{6} \varepsilon + \cdots\right)
\left(\frac{\alpha}{4\pi\varepsilon}\right)^2\,,
\end{equation*}
we finally obtain
\begin{equation}
\zeta_\psi(M) = \frac{Z'_\psi(\alpha',a')}{Z_\psi(\alpha,a)} \zeta_\psi^0
= 1 - \frac{5}{6} \left(\frac{\alpha(M)}{4\pi}\right)^2 + \cdots
\label{electron:zetares}
\end{equation}

The RG equation
\begin{equation}
\frac{d\log\zeta_\psi(\mu)}{d\log\mu}
+ \gamma_\psi(\alpha(\mu),a(\mu)) - \gamma'_\psi(\alpha'(\mu),a'(\mu)) = 0
\label{electron:RG}
\end{equation}
can be used to find $\zeta_\psi(\mu)$ for $\mu\neq M$.
In contrast to the case of $\zeta_A(\mu)$~(\ref{photon:RG}),
now $\gamma_\psi-\gamma_\psi'$ is of order $\alpha^2$,
so that changes of $\mu$ of order $\alpha$
(such as, e.g., (\ref{Decoupling:Mmu}))
don't change the coefficient of $\alpha^2$ in~(\ref{electron:zetares}).

\subsection{Decoupling: electron charge}
\label{S:charge}

\begin{figure}[ht]
\begin{center}
\begin{picture}(52,28)
\put(26,14){\makebox(0,0){\includegraphics{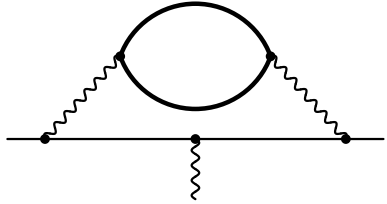}}}
\end{picture}
\end{center}
\caption{Two-loop diagram for $\Lambda(0,0)$}
\label{F:muon}
\end{figure}

The proper vertex $e_0\Gamma$ with the external propagators attached
(two electron propagators $S$ and one photon propagator $D$)
is the Green function of the fields $\bar{\psi}_0$, $\psi_0$, $A_0$
(i.e., the Fourier transform of the vacuum average
of the $T$-product of these three fields).
Therefore, the relation between this quantity in the full theory
and in the low-energy effective theory is
\begin{equation}
e_0 \Gamma S S D = \zeta_\psi^0 \left(\zeta_A^0\right)^{1/2}
e_0' \Gamma' S' S' D'\,,
\label{charge:Gamma1}
\end{equation}
or, taking into account $S=\zeta_\psi^0 S'$, $D=\zeta_A^0 D'$,
\begin{equation}
e_0 \Gamma^\mu = \left(\zeta_\psi^0\right)^{-1} \left(\zeta_A^0\right)^{-1/2}
e_0' \Gamma^{\prime\mu}
\label{charge:Gamma2}
\end{equation}
The vertex $\Gamma^\mu(p,p')=\gamma^\mu+\Lambda^\mu(p,p')$ on the mass shell
has two $\gamma$-matrix structures when sandwiched between physical spinors;
when $q=p'-p=0$, only one of them remains:
\begin{equation}
\Gamma^\mu = Z_\Gamma^{\text{os}} \gamma^\mu\,.
\label{charge:full}
\end{equation}
Only diagrams with muon loops contribute to $\Lambda^\mu(p,p)$,
all the other diagrams contain no scale;
such diagrams first appear at two loops (Fig.~\ref{F:muon}).
In the effective theory $Z_\Gamma^{\prime\text{os}}=1$,
because all diagrams for $\Lambda^{\prime\mu}(p,p)$ vanish.
Therefore,
\begin{equation}
\Gamma^\mu = \zeta_\Gamma^0 \Gamma^{\prime\mu}\,,\quad
\zeta_\Gamma^0 = \frac{Z_\Gamma^{\text{os}}}{Z_\Gamma^{\prime\text{os}}}
= Z_\Gamma^{\text{os}}\,,
\label{charge:zetaGamma0}
\end{equation}
and we obtain from~(\ref{charge:Gamma2})
\begin{equation}
\zeta_\alpha^0 = \left(\zeta_\Gamma^0 \zeta_\psi^0\right)^{-2}
\left(\zeta_A^0\right)^{-1}\,.
\label{charge:zeta0}
\end{equation}

The situation in QED is simpler, due to the Ward identity
\begin{equation}
\Gamma^\mu(p,p) = \frac{\partial S^{-1}(p)}{\partial p_\mu}\,.
\label{charge:Ward}
\end{equation}
Near the mass shell $p^2=0$,
\begin{equation}
S(p) = \frac{Z_\psi^{\text{os}}}{\rlap/p}\,,
\label{charge:S}
\end{equation}
and therefore
\begin{equation}
Z_\Gamma^{\text{os}} Z_\psi^{\text{os}} = 1\,.
\label{charge:Wardos}
\end{equation}
Similarly, in the effective theory, $Z_\Gamma^{\prime\text{os}}Z_\psi^{\prime\text{os}}=1$
(in fact, these two renormalization constants are equal to 1 separately).
Therefore,
\begin{equation}
\zeta_\Gamma^0 \zeta_\psi^0 = 1\,,\quad
\zeta_\alpha^0 = \left(\zeta_A^0\right)^{-1}\,.
\label{charge:Ward0}
\end{equation}

The bare propagator and vertex are related to the \MS{} renormalized ones
by $S(p) = Z_\psi S_r(p)$, $\Gamma^\mu = Z_\Gamma \Gamma^\mu_r$,
where $Z_\psi$, $Z_\Gamma$ are minimal renormalization constants,
and $S_r$, $\Gamma^\mu_r$ are finite at $\varepsilon\to0$.
The Ward identity~(\ref{charge:Ward}) implies that $Z_\Gamma Z_\psi$
is finite at $\varepsilon\to0$, but the only minimal renormalization constant
with this property is 1:
\begin{equation}
Z_\Gamma Z_\psi = 1\,.
\label{charge:WardMS}
\end{equation}
Therefore, $Z_\alpha=(Z_\Gamma Z_\psi)^{-2} Z_A^{-1}=Z_A^{-1}$;
similarly, $Z'_\alpha=Z_A^{\prime-1}$, and we obtain
\begin{equation}
\zeta_\alpha = \zeta_\alpha^0 \frac{Z_\alpha'}{Z_\alpha}
= \left(\zeta_A^0 \frac{Z_A'}{Z_A}\right)^{-1}
= \zeta_A^{-1}\,.
\label{charge:Ward2}
\end{equation}

This means that the running charge in full QED (with both electrons and muons)
at $\mu=\mu_0$~(\ref{Decoupling:mu0}) is slightly larger than in the
low-energy effective QED (with only electrons):
\begin{equation}
\alpha(\mu_0) = \zeta_\alpha(\mu_0) \alpha'(\mu_0)\,,\quad
\zeta_\alpha(\mu_0) = 1 + \frac{13}{3} \left(\frac{\alpha(\mu_0)}{4\pi}\right)^2 + \cdots
\label{charge:zeta2}
\end{equation}
(see~(\ref{Decoupling:zeta2a}));
the same is true at $\mu=M_{\text{os}}$~(\ref{Decoupling:zeta2b}).

\subsection{Decoupling: bilinear electron currents}
\label{S:j}

Various operators in the full theory can be expressed via operators
in the low energy effective theory.
All operators with appropriate quantum numbers appear;
contributions of higher-dimensional operators are suppressed
by powers of $1/M$.
This $1/M$ expansion of a full-theory operator means
that its on-shell matrix elements with light external particles
having small momenta $p_i$ (and physical polarizations),
being expanded in $p_i/M$ up to some finite order,
coincides with the $1/M$ series in matrix elements
calculated in the effective theory.
As an example, here we shall discuss
flavour-nonsinglet bilinear light-fermion currents~(\ref{App:j0}).
The full-theory renormalized current is related
to the corresponding effective-theory current by
\begin{equation}
j_n(\mu) = \zeta_{jn}(\mu) j'_n(\mu)+ \mathcal{O}(1/M^2)\,.
\label{j:zetadef}
\end{equation}

The on-shell matrix element of $j_n(\mu)$ is
$M_n(p,p';\mu) = Z_\psi^{\text{os}} Z_{jn}^{-1}(\alpha(\mu)) \Gamma_n(p,p')$.
It should be equal to $\zeta_{jn}(\mu) M'_n(p,p';\mu)$,
where $M'_n(p,p';\mu) =
Z_\psi^{\prime\text{os}} Z_{jn}^{\prime-1}(\alpha'(\mu)) \Gamma'_n(p,p')$.
Both matrix elements are UV-finite;
their IR divergences coincide,
because both theories are identical in the IR region.
Any on-shell momenta $p$, $p'$ can be used;
it is easiest to set $p=p'=0$, thus excluding power-suppressed terms.
The proper vertex $\Gamma_n(p,p')=\gamma_{(n)}+\Lambda_n(p,p')$ at $p=p'=0$
has the structure $\Gamma_n(0,0)=\gamma_{(n)}\Gamma_n$,
where $\gamma_{(n)}$ is the Dirac matrix~(\ref{App:gamman}),
and $\Gamma_n=1+\Lambda_n$ is scalar.
Therefore,
\begin{equation}
\zeta_{jn}(\mu) = \frac{Z'_{jn}(\alpha'(\mu))}{Z_{jn}(\alpha(\mu))}
\zeta_{jn}^0\,,\quad
\zeta_{jn}^0 = Z_\psi^{\text{os}} \Gamma_n\,,
\label{j:zeta}
\end{equation}
because in the effective theory $Z_\psi^{\prime\text{os}}=1$, $\Gamma'_{jn}=1$.
This can also be understood by comparing Green functions
of $\bar{\psi}_0$, $\psi_0$, and $j_{n0}$, similarly to Sect.~\ref{S:charge}.

\begin{figure}[b]
\begin{center}
\begin{picture}(52,28)
\put(26,14){\makebox(0,0){\includegraphics{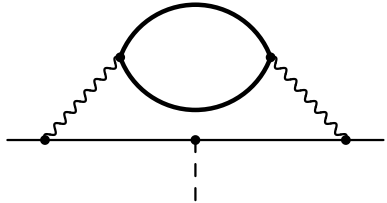}}}
\end{picture}
\end{center}
\caption{Two-loop on-shell matrix element of a QCD bilinear quark current}
\label{F:decj}
\end{figure}

Only diagrams with muon loops contribute to $\Lambda_n(0,0)$;
the only relevant two-loop diagram is shown in Fig.~\ref{F:decj}.
It is equal to
\begin{equation*}
\Lambda_n(0,0) = -i e_0^2 \int \frac{d^d k}{(2\pi)^d}
\frac{\gamma^\mu\rlap/k\gamma_{(n)}\rlap/k\gamma^\nu(k^2 g_{\mu\nu}-k_\mu k_\nu)
\Pi(k^2)}{(k^2)^4}\,,
\end{equation*}
where $\Pi(k^2)$ is the muon loop contribution to the photon self-energy.
Averaging over the directions of $k$ and using~(\ref{App:h}) we obtain
\begin{equation*}
\Lambda_n = -i e_0^2 \left(\frac{(d-2n)^2}{d}-1\right)
\int \frac{d^d k}{(2\pi)^d} \frac{\Pi(k^2)}{(-k^2)^2}\,.
\end{equation*}
Using~(\ref{electron:zeta0res}) and the integral~(\ref{electron:I}) we obtain
\begin{equation}
\zeta_{jn}^0 = 1
- \frac{e_0^4 M_0^{-4\varepsilon}}{(4\pi)^d} \Gamma^2(\varepsilon)
\frac{8 (d-6) (n-1) (n-d+1)}{d (d-2) (d-5) (d-7)}\,.
\label{j:zeta0}
\end{equation}
The ratio $Z_{jn}/Z'_{jn}$ is given by the formula
similar to~(\ref{electron:Zratio}):
\begin{equation*}
\frac{Z_{jn}}{Z'_{jn}} = 1
+ \frac{1}{9} (n-1) \left[6(n-3)-(n-15)\varepsilon\right]
\left(\frac{\alpha}{4\pi\varepsilon}\right)^2\,,
\end{equation*}
and we arrive at~\cite{G:98}
\begin{equation}
\zeta_{jn}(M) = 1 + \frac{1}{54} (n-1) (85n-267)
\left(\frac{\alpha(M)}{4\pi}\right)^2 + \cdots
\label{j:zeta2}
\end{equation}
The three-loop correction has been calculated in~\cite{GSS:06}.

The vector current has $\zeta_{j1}=1$ to all orders.
For the vector current with a diagonal flavour matrix $\tau$,
the integral of its 0th component is an integer ---
the difference between the numbers of light leptons and antileptons
weighted by the diagonal elements of $\tau$.
This difference is the same in the full QED
and in the low-energy effective theory.
The same holds for non-diagonal $\tau$ by flavour symmetry.
We can also see this explicitly.
Multiplying the Ward identity
\begin{equation*}
\Gamma_1^\mu(0,0)
= \gamma^\mu - \left.\frac{\partial\Sigma(p)}{\partial p_\mu}\right|_{p=0}
= \gamma^\mu \left(1 - \Sigma_V(0)\right)
\end{equation*}
by $Z_\psi^{\text{os}}=\left[1-\Sigma_V(0)\right]^{-1}$,
we obtain just $\Gamma_1=1$.
Taking account of the fact that the vector current does not renormalize
($Z_{j1}=1$, $Z'_{j1}=1$) yields $\zeta_{j1}(\mu)=1$.

The currents $j_4$ and $j_3$ differ from $j_0$ and $j_1$
by insertion of $\gamma_5^{\text{HV}}$.
They are related to those containing $\gamma_5^{\text{AC}}$
by~(\ref{App:ZPdef}) and~(\ref{App:ZAdef}).
Inserting $\gamma_5^{\text{AC}}$ does not change the decoupling coefficient.
Therefore,
\begin{equation}
\zeta_{j4} = \zeta_{j0} \frac{Z'_P}{Z_P}\,,\quad
\zeta_{j3} = \zeta_{j1} \frac{Z'_A}{Z_A}\,,
\label{j:g5}
\end{equation}
where $Z_{P,A}$ are given by~(\ref{App:ZPA2}) at two loops,
and $Z'_{P,A}$ contain $n_l$ instead of $n_f=n_l+1$.

The $1/M^2$ correction to the full-theory vector current $j_1$
(see~(\ref{j:zetadef})) has been calculated in~\cite{C:93}.
Only one dimension-5 effective-theory operator
appears in this correction, namely, $\partial^2 j'_1$.
The coefficient of this operator contains $\log(\mu^2/M^2)$.
The leading effective-theory current $j'_1$ also depends on $\mu$
when we take into account the leading term $c_1 O_1$ in $L_1$~(\ref{Cont:L});
its derivative in $\log\mu$ is proportional to $\partial^2 j'_1$.
Of course, these two sources of $\mu$-dependence cancel,
and the full-theory vector current $j_1$ does not depend on $\mu$
also at the $1/M^2$ level.

\subsection{Decoupling: electron mass}
\label{S:mass}

In the previous Sections, we considered QED with massless electrons
and heavy muons (with mass $M$).
Now let's take the electron mass into account as a small correction.
We shall expand everything up to linear terms in $m$.
The electron propagator in the full theory is given by~(\ref{App:S});
in the low-energy effective theory,
it involves $\Sigma'_{V,S}$ instead of $\Sigma_{V,S}$.
These two propagators are related by $\zeta^0_\psi$:
\begin{equation}
\frac{1}{1-\Sigma_V(p^2)} \frac{1}{\displaystyle
\rlap/p - \frac{1+\Sigma_S(p^2)}{1-\Sigma_V(p^2)} m_0}
= \zeta^0_\psi \frac{1}{1-\Sigma'_V(p^2)} \frac{1}{\displaystyle
\rlap/p - \frac{1+\Sigma'_S(p^2)}{1-\Sigma'_V(p^2)} m'_0}\,.
\label{mass:S}
\end{equation}
Comparing the overall factors, we recover~(\ref{electron:zeta0}).
The denominators should also coincide:
\begin{equation}
\frac{1+\Sigma_S(p^2)}{1-\Sigma_V(p^2)} m_0 =
\frac{1+\Sigma'_S(p^2)}{1-\Sigma'_V(p^2)} m'_0\,,
\label{mass:mm}
\end{equation}
This means that the on-shell masses are the same in both theories
($m_{\text{os}}=m'_{\text{os}}$)
--- they can be measured at large distances, by a macroscopic experiment
(weighting $N$ electrons)%
\footnote{This can also be written as
$\zeta_m^0 = Z_m^{\text{os}}/Z_m^{\prime\text{os}}$
where $m_0=Z_m^{\text{os}}m_{\text{os}}$,
$m'_0=Z_m^{\prime\text{os}}m'_{\text{os}}$.}:
\begin{equation}
\zeta_m^0 = \frac{m_0}{m'_0} = \left(\zeta_\psi^0\right)^{-1}
\frac{1+\Sigma'_S(p^2)}{1+\Sigma_S(p^2)}\,.
\label{mass:zeta0}
\end{equation}
This equation should hold for all $m\ll M$, $p\ll M$.
It is easiest to set $m=0$, and use~(\ref{App:Sigma}).
Then, setting $p=0$, we obtain
\begin{equation}
\zeta_m^0 = \left(\zeta_{j0}^0\right)^{-1}\,.
\label{mass:zetam0}
\end{equation}
Recalling~(\ref{App:Zm}), we finally arrive at
\begin{equation}
\zeta_m(\mu) = \frac{Z_m'}{Z_m} \zeta_m^0 = \zeta_{j0}^{-1}(\mu)\,.
\label{mass:zetam}
\end{equation}
In other words, $m(\mu)[\bar{\psi}\tau\psi]_\mu
=m'(\mu')[\bar{\psi}\tau\psi]'_{\mu'}$
does not vary with $\mu$ and $\mu'$,
and does not change when one goes from the full theory to the low-energy one.
The \MS{} mass decoupling constant is
\begin{equation}
\zeta_m(M) = 1 - \frac{89}{18} \left(\frac{\alpha(M)}{4\pi}\right)^2
+ \cdots
\label{mass:zetam2}
\end{equation}
its $\mu$ dependence can be found from the RG equation
\begin{equation}
\frac{d\log\zeta_m(\mu)}{d\log\mu}
+ \gamma_m(\alpha(\mu)) - \gamma'_m(\alpha'(\mu)) = 0
\label{mass:RG}
\end{equation}
(the difference $\gamma_m-\gamma_m'$ is of order $\alpha^2$).

\subsection{Electron magnetic moment}
\label{S:mag}

When the electron mass $m\neq0$, the electron helicity is no longer conserved.
Therefore, the effective Lagrangian contains the contribution
\begin{equation}
L_m = - \frac{1}{4} c_m O_m\,,\quad
O_m = m e \bar{\psi} F_{\mu\nu} \sigma^{\mu\nu} \psi
\label{mag:Om}
\end{equation}
of the magnetic operator $O_m$ having dimension 6.

Let's discuss scattering of an on-shell electron (with a physical polarization)
in an external magnetic field in full QED
(we'll return to the low-energy effective theory later).
It is described by two form factors $F_{1,2}(q^2)$.
The anomalous magnetic moment is given by $\mu=F_2(0)$.
In order to find it, we need to expand the vertex function in $q$
up to the linear term:
\begin{equation}
\raisebox{-10mm}{\begin{picture}(29,22)
\put(14.5,12.5){\makebox(0,0){\includegraphics{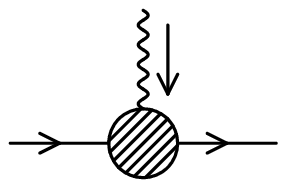}}}
\put(6,0){\makebox(0,0)[b]{$mv$}}
\put(19.5,16){\makebox(0,0)[l]{$q$}}
\end{picture}}
= i e_0 \left[ \Gamma_0^\mu + \Gamma_1^{\mu\nu} \frac{q_\nu}{m} + \cdots \right]\,.
\label{mag:Gamma}
\end{equation}
Then (see, e.\,g., \cite{G:07})
\begin{equation}
\begin{split}
\mu = \frac{Z_\psi^{\text{os}}}{d-2}
\biggl[& \frac{1}{4} \Tr (\gamma_\mu-dv_\mu) \Gamma_0^\mu (\rlap/v+1)\\
&{} + \frac{2}{d-1} \frac{1}{4} \Tr
\left( \gamma_\mu \gamma_\nu + \gamma_\mu v_\nu - \gamma_\nu v_\mu - v_\mu v_\nu \right)
\Gamma_1^{\mu\nu} (\rlap/v+1) \biggr]\,.
\end{split}
\label{mag:F20}
\end{equation}
As expected, at the tree level ($\Gamma_0^\mu=\gamma^\mu$, $\Gamma_1^{\mu\nu}=0$)
the anomalous magnetic moment is $\mu=0$.

\begin{figure}[ht]
\begin{center}
\begin{picture}(31,20)
\put(15.5,10){\makebox(0,0){\includegraphics{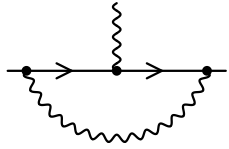}}}
\end{picture}
\end{center}
\caption{The one-loop contribution to the electron magnetic moment}
\label{F:mag0}
\end{figure}

The first non-vanishing contribution appears at one loop (Fig.~\ref{F:mag0}):
\begin{equation}
\mu = \frac{e_0^2 m^{-2\varepsilon}}{(4\pi)^{d/2}} \Gamma(\varepsilon)
\frac{(d-4)(d-5)}{d-3} + \cdots
\Rightarrow 2 \frac{\alpha}{4\pi} + \cdots
\label{mag:mu0}
\end{equation}
where dots mean higher-loop contributions.

\begin{figure}[ht]
\begin{center}
\begin{picture}(31,23.5)
\put(15.5,11.75){\makebox(0,0){\includegraphics{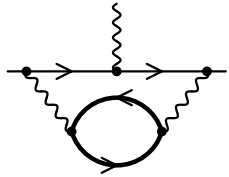}}}
\end{picture}
\end{center}
\caption{The muon-loop contribution to the electron magnetic moment}
\label{F:mag}
\end{figure}

There are contributions with muon loops.
They first appear at two loops (Fig.~\ref{F:mag}).
We shall consider this diagram using the method of regions,
see the textbook~\cite{S:02}.
In fact, in simple situations (like the large-mass expansion)
this method is known for a long time, see, e.\,g., \cite{GL:87}.

Using the projector~(\ref{mag:F20}) we obtain
\begin{align}
\mu =& \frac{i e_0^2}{d-1} \int \frac{d^d k}{(2\pi)^d} \Pi(k^2)
\Biggl[ \frac{1}{m^2} \left( \frac{(d+1)(d-2)}{2 D_2} - \frac{d^2-d-3}{D_1}
+ \frac{(d+2)(d-3)D_2}{2 D_1^2} + \frac{D_2^2}{D_1^3} \right)
\nonumber\\
&{} + \frac{2}{d-2} \left( - \frac{d^2-4d+5}{D_1 D_2} + \frac{2d^2-9d+13}{D_1^2}
+ \frac{2(d-3)D_2}{D_1^3} \right) - \frac{16 m^2}{(d-2)D_1^3} \Biggr]\,,
\label{mag:gen}
\end{align}
where the denominators are
\begin{equation*}
D_1 = m^2 - (k+mv)^2\,,\quad
D_2 = - k^2\,,
\end{equation*}
and $\Pi(k^2)$ is the muon-loop contribution to the photon self-energy.

There are two integration regions in the diagram Fig.~\ref{F:mag}:
\begin{itemize}
\item The muon loop is hard (momenta $\sim M$),
the other loop is soft (momenta $\sim m$);
\item both loops are hard.
\end{itemize}
In the first region, we can expand the muon loop $\Pi(k^2)$
in the small momentum $k$.
By dimensionality, these one-loop terms with a single scale $M_0$
have the form $e_0^2 M_0^{-2\varepsilon} (k^2/M_0^2)^n$, see~(\ref{Coulomb:Pi1}).
When we substitute this expansion into the soft loop integral
(with a single scale $m$), these terms produce
$e_0^4 M_0^{-2\varepsilon} m^{-2\varepsilon} (m^2/M_0^2)^n$.
Adding this contribution to that of Fig.~\ref{F:mag0}~(\ref{mag:mu0})%
\footnote{It can be reproduced from~(\ref{mag:gen})
if we formally put $\Pi(k^2)=1$,
because the photon propagator with this insertion becomes simply
the Landau-gauge free propagator.},
we have
\begin{equation}
\mu = \frac{e_0^2 m^{-2\varepsilon}}{(4\pi)^{d/2}} \Gamma(\varepsilon) \frac{(d-4)(d-5)}{d-3}
\left[ 1 - \frac{4}{3} \frac{e_0^2 M_0^{-2\varepsilon}}{(4\pi)^{d/2}} \Gamma(\varepsilon)
\left( 1 - \frac{(d-3)(d-4)}{5(d-5)} \frac{m^2}{M_0^2} \right) \right] + \cdots
\label{mag:soft0}
\end{equation}
where dots mean all the other contributions.

In the purely hard region, the electron lines carry large momenta $\sim M$,
and we can expand their propagators in $m$ using $D_1=D_2-2mv\cdot k$.
This is a regular expansion (only integer powers of $m^2$ appear);
by dimensionality, the two-loop integrals (having a single scale $M_0$)
produce $e_0^4 M_0^{-4\varepsilon} (m^2/M_0^2)^n$.
Averaging over $k$ directions we see that the first non-vanishing contribution
appears at $\mathcal{O}(m^2)$:
\begin{equation}
\begin{split}
\mu_h &= - 32 \frac{m^2}{M_0^2} \frac{e_0^4 M_0^{-4\varepsilon}}{(4\pi)^d}
\frac{d-4}{d(d+2)} I_3\\
&= - 32 \frac{m^2}{M_0^2} \frac{e_0^4 M_0^{-4\varepsilon}}{(4\pi)^d} \Gamma^2(\varepsilon)
\frac{(d-4)^2 (d-8)}{d(d+2)(d-2)(d-5)(d-7)(d-9)}\\
&\Rightarrow - \frac{32}{45} \frac{m^2}{M^2}
\left(\frac{\alpha}{4\pi}\right)^2\,,
\end{split}
\label{mag:hard}
\end{equation}
where $I_3$ is defined in~(\ref{electron:Idef}).

How can we understand this in the framework of the low-energy effective theory?
The one-loop contribution (Fig.~\ref{F:mag0}) is given
by the formula~(\ref{mag:mu0}) with $e_0^2\to e_0^{\prime2}$.
The leading correction in~(\ref{mag:soft0}) transforms the full-QED charge $e_0^2$
in the leading term into the effective-theory one $e_0^{\prime2}=(\zeta_\alpha^0)^{-1}e_0^2$.
The contribution of the $\mathcal{O}(k^2)$ term in $\Pi(k^2)$ in~(\ref{mag:soft0}),
\begin{equation}
\mu_s = \frac{4}{15} \frac{m^2}{M_0^2}
\frac{e_0^4 m^{-2\varepsilon} M_0^{-2\varepsilon}}{(4\pi)^d}
\Gamma^2(\varepsilon) (d-4)^2
\Rightarrow \frac{16}{15} \frac{m^2}{M^2}
\left(\frac{\alpha}{4\pi}\right)^2\,,
\label{mag:soft}
\end{equation}
corresponds to the contact-interaction diagram
(the second diagram in Fig.~\ref{F:Eva2}) in the effective theory
(the first diagram in Fig.~\ref{F:Eva2}
does not contribute to the anomalous magnetic moment,
because it has the pure $\gamma^\mu$ Dirac structure).
Finally, the hard contribution~(\ref{mag:hard})
is produced by the local vertex of the operator~(\ref{mag:Om})
in the effective theory: $\mu_h = c_m m^2$.
Now we see why $\mu_h$ starts from $\mathcal{O}(m^2)$.

\begin{figure}[ht]
\begin{center}
\begin{picture}(88,19)
\put(13,11){\makebox(0,0){\includegraphics{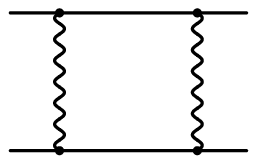}}}
\put(49,11){\makebox(0,0){\includegraphics{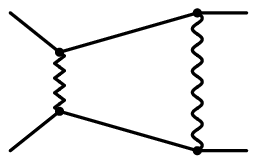}}}
\put(80,11){\makebox(0,0){\includegraphics{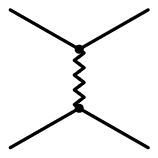}}}
\put(13,0){\makebox(0,0)[b]{$a$}}
\put(49,0){\makebox(0,0)[b]{$b$}}
\put(80,0){\makebox(0,0)[b]{$c$}}
\end{picture}
\end{center}
\caption{Contributions to electron--electron scattering}
\label{F:ee}
\end{figure}

This correspondence between contributions of various integration regions
in diagrams of the full theory
(which contains several widely separated energy scales)
and the low-energy effective theory
(which contains higher-dimensional local operators in its Lagrangian)
is quite general.
Let's briefly discuss one more example ---
electron--electron scattering at small external momenta $p_i\sim m$.
In the full theory there are diagrams with muon loop(s).
For example, let's consider the diagram of Fig.~\ref{F:O3}.
It contains two integration regions:
(1) the muon loop is hard, the other one is soft;
(2) both loops are hard.
In the first region, we expand the muon loop in $k^2$
($e_0^2 M_0^{-2\varepsilon} (k^2/M_0^2)^n$);
the remaining soft loop is a complicated non-analytical function
of the soft variables: $m$ and $p_i$.
The $\mathcal{O}(1)$ term in $\Pi(k^2)$ just replaces
$e_0^2\to e_0^{\prime2}$ in the lower-order diagram (Fig.~\ref{F:ee}$a$).
The $\mathcal{O}(k^2)$ term corresponds to the effective-theory diagram
with the $\mathcal{O}(\alpha^2)$ term in the contact interaction $c_1$
(Fig.~\ref{F:ee}$b$).
In the purely hard contribution, the integrand can be expanded
in the small variables $m$ and $p_i$.
Truncating this regular expansion, we obtain a polynomial in $p_i$
which gives a local vertex (produced by a local operator
in the effective Lagrangian), Fig.~\ref{F:ee}$c$.
In particular, the zeroth term of this expansion
produces $\mathcal{O}(\alpha^3)$ contributions
to the Wilson coefficients $c_1$ and $c_3$.

\section{Decoupling of heavy flavours in QCD}
\label{S:QCD}

\subsection{Effective low-energy QCD}
\label{S:EQCD}

In QED, effects of decoupling of muon loops are tiny.
Also, pion pairs become important at about the same energies
as muon pairs, so that QED with electrons and muons
is a model with a narrow region of applicability.
Therefore, everything we discussed in Sect.~\ref{S:mu}
is not particularly important, from the practical point of view.

In QCD, decoupling of heavy flavours is fundamental and omnipresent.
It would be a huge mistake to use the full 6-flavour QCD
at characteristic energies of a few GeV, or a few tens of GeV:
running of $\alpha_s(\mu)$ and other quantities would be grossly inadequate,
convergence of perturbative series would be awful because of large logarithms.
In most cases, anybody working in QCD uses an effective low-energy QCD,
where a few heaviest flavours have been removed.
Therefore, it is important to understand decoupling in QCD.
And to this end the lessons of QED are very helpful.

Let's consider QCD with a single heavy flavour having mass $M$;
for simplicity, all other flavours are supposed to be massless.
Then the behaviour of light quarks and gluons at low momenta $p_i\ll M$
is described by the low-energy effective theory.
Its Lagrangian is the usual QCD Lagrangian
(of course, without the heavy-quark field)
plus higher-dimensional terms
(whose coefficients are suppressed by powers of $1/M$).
Power corrections to the Lagrangian first appear at dimension 6.

Unlike the QED case~(\ref{Photonia:O6}),
there exists a non-vanishing dimension-6 operator
constructed from 3 $G^a_{\mu\nu}$:
\begin{equation}
O_{g1}^0 = g_0 f^{abc}
G_{0\lambda}^a{}^\mu G_{0\mu}^b{}^\nu G_{0\nu}^c{}^\lambda
\label{EQCD:O1}
\end{equation}
(a similar operator with $d^{abc}$ is identically zero, as in QED~(\ref{Photonia:O6})).
One-loop renormalization of the operator~(\ref{EQCD:O1})
has been considered in~\cite{NT:83,M:84}.
There is also the operator
\begin{equation}
O_{g2}^0 = (D^\mu G^a_{0\lambda\mu}) (D_\nu G_0^{a\lambda\nu})
\label{EQCD:O2}
\end{equation}
which reduces to a quark operator due to equations of motion.
The operator $G^a_{0\mu\nu} D^2 G_0^{a\mu\nu}$ reduces to~(\ref{EQCD:O2}),
similarly to the QED case (Sect.~\ref{S:Photonia}).

The local four-quark dimension-6 operators are
\begin{equation}
\begin{split}
O_{qn}^0 &= \biggl(\sum_q \bar{q}_0 \gamma_{(n)} q_0\bigg)
\biggl(\sum_q \bar{q}_0 \gamma_{(n)} q_0\biggr)\,,\\
\tilde{O}_{qn}^0 &= \biggl(\sum_q \bar{q}_0 \gamma_{(n)} t^a q_0\biggr)
\biggl(\sum_q \bar{q}_0 \gamma_{(n)} t^a q_0\biggr)\,.
\end{split}
\label{EQCD:Oq}
\end{equation}
Only operators with odd $n$ conserve the light-quark helicity.
Those with even $n$ can only appear in the Lagrangian
being multiplied by a power of light-quark masses;
these terms have higher dimensions, and we don't consider them here.
Operators~(\ref{EQCD:Oq}) with $n\ge5$ are evanescent,
and are eliminated by renormalization (Sect.~\ref{S:4f}).
Only 4 operators $O_{q1}$, $O_{q3}$, $\tilde{O}_{q1}$, $\tilde{O}_{q3}$
appear in the effective QCD Lagrangian (at the dimension-6 level).
The operator $O^0_{g2} - g_0^2 \tilde{O}^0_{q1}$
vanishes due to equations of motion.
Renormalization of the four-quark operators~(\ref{EQCD:Oq})
can be considered similarly to Sect.~\ref{S:4f};
the only extra complication is their colour structures.

The $\mathcal{O}(q^2)$ term in the one-loop gluon self-energy~(\ref{Coulomb:Pi})
(which is given by~(\ref{Coulomb:Pi1}) with $e_0^2\to g_0^2$ and the colour factor $T_F$)
gives the term $c^0_{g2} O^0_{g2}$ in the Lagrangian with
\begin{equation}
c_{g2}(M) = - \frac{2}{15} \frac{T_F}{M^2}
\left( \frac{\alpha_s(M)}{4\pi} + \mathcal{O}(\alpha_s^2(M)) \right)
\label{EQCD:cg2}
\end{equation}
(of course, our one-loop calculation leaves higher orders in $\alpha_s$ unknown;
if we choose $\mu\sim M$, there are no large logarithms,
and the error due to truncation of the perturbative series is minimal).
We are interested only in the $S$-matrix;
therefore, we can eliminate this term in favour of $\tilde{c}^0_{q1} \tilde{O}^0_{q1}$
with
\begin{equation}
\tilde{c}_{q1}(M) = - \frac{2}{15} \frac{T_F}{M^2}
\left( \alpha_s^2(M) + \mathcal{O}(\alpha_s^3(M)) \right)\,.
\label{EQCD:cq1}
\end{equation}

\begin{figure}[ht]
\begin{center}
\begin{picture}(36,32)
\put(18,17){\makebox(0,0){\includegraphics{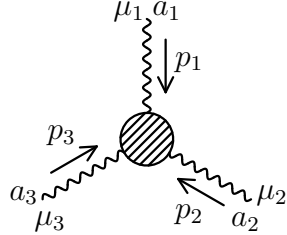}}}
\put(15.5,30){\makebox(0,0){$\mu_1$}}
\put(20.5,30){\makebox(0,0){$a_1$}}
\put(23.5,23.25){\makebox(0,0){$p_1$}}
\put(34.5,5.5){\makebox(0,0){$\mu_2$}}
\put(31,2){\makebox(0,0){$a_2$}}
\put(5,2){\makebox(0,0){$\mu_3$}}
\put(1.5,5.5){\makebox(0,0){$a_3$}}
\put(23.5,3){\makebox(0,0){$p_2$}}
\put(6.5,13.5){\makebox(0,0){$p_3$}}
\end{picture}
\end{center}
\caption{Three-gluon scattering amplitude}
\label{F:G3}
\end{figure}

The 3-gluon interaction $c^0_{g1} O^0_{g1}$ also first appears at one loop.
The coefficient $c^0_{g1}$ can be found by matching
the 3-gluon scattering amplitudes (Fig.~\ref{F:G3}, $p_1+p_2+p_3=0$)
in the full theory and in the effective one.
Diagrams without heavy-quark loops are the same in both theories.
We are interested only in diagrams with (at least one) heavy-quark loop
in the full QCD.
Contributions of the fully hard regions in such diagrams
can be expanded in the small momenta $p_i$ up to cubic terms.
We cannot put all 3 legs on-shell
(as we did in the case of the 4-photon scattering amplitude in Sect.~\ref{S:Qedland}).
Therefore, when doing the effective-theory calculation,
we have to take both operators~(\ref{EQCD:O1}), (\ref{EQCD:O2}) into account.

\begin{figure}[t]
\begin{center}
\begin{picture}(36,33)
\put(18,18){\makebox(0,0){\includegraphics{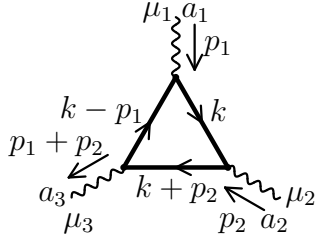}}}
\put(15.5,31){\makebox(0,0){$\mu_1$}}
\put(20.5,31){\makebox(0,0){$a_1$}}
\put(23.5,26.5){\makebox(0,0){$p_1$}}
\put(34.5,6.5){\makebox(0,0){$\mu_2$}}
\put(31,3){\makebox(0,0){$a_2$}}
\put(5,3){\makebox(0,0){$\mu_3$}}
\put(1.5,6.5){\makebox(0,0){$a_3$}}
\put(25.5,3){\makebox(0,0){$p_2$}}
\put(2,13.5){\makebox(0,0){$p_1+p_2$}}
\put(18,7.5){\makebox(0,0){$k+p_2$}}
\put(23.5,18){\makebox(0,0){$k$}}
\put(8,18){\makebox(0,0){$k-p_1$}}
\end{picture}
\end{center}
\caption{Heavy-quark loop contribution to the 3-gluon scattering amplitude}
\label{F:Gq}
\end{figure}

The scattering amplitude appears at the tree level ---
just the 3-gluon vertex.
It should be multiplied by the one-loop
external-legs renormalization factor $Z_A^{3/2}$.
All contributions to $Z_A$ without a heavy-quark loop
appear also in the effective theory and hence cancel.
Leaving only the heavy-quark contribution to $Z_A$ we get
\begin{equation}
- g_0 f^{a_1 a_2 a_3} V^{\mu_1 \mu_2 \mu_3}
\left[ 1
- 2 T_F \frac{g_0^2 M_0^{-2\varepsilon}}{(4\pi)^{d/2}} \Gamma(\varepsilon)
\right]\,,
\label{EQCD:0}
\end{equation}
where the 3-gluon vertex tensor structure $V^{\mu_1 \mu_2 \mu_3}$
is linear in the external momenta $p_i$.
There are also one-loop vertex diagrams with a heavy-quark loop:
the diagram in Fig.~\ref{F:Gq} and the one with the opposite direction
of the quark line.
The $d^{a_1 a_2 a_3}$ contributions cancel (just like in QED)
and the $f^{a_1 a_2 a_3}$ ones are equal
(so that we can calculate the $f^{a_1 a_2 a_3}$ term
of the diagram in Fig.~\ref{F:Gq} and then double it).
We expand this diagram in the external momenta $p_i$.
Only odd powers appear; we need terms up to cubic ones.
Averaging the integrand over $k$ directions and using~(\ref{Qedland:V1})
we obtain
\begin{equation}
T_F f^{a_1 a_2 a_3} \frac{g_0^2 M_0^{-2\varepsilon}}{(4\pi)^{d/2}} \Gamma(\varepsilon)
\left[ - \frac{4}{3} g_0 V^{\mu_1 \mu_2 \mu_3}
+ i \frac{d-4}{180} (T_1^{\mu_1 \mu_2 \mu_3} + 12 T_2^{\mu_1 \mu_2 \mu_3})
+ \mathcal{O}(p_i^5) \right]\,,
\label{EQCD:1}
\end{equation}
where the 3-gluon matrix elements $f^{a_1 a_2 a_3} T_{1,2}^{\mu_1 \mu_2 \mu_3}$
of the operators $O^0_{1,2}$ are cubic in $p_i$.
Adding~(\ref{EQCD:0}) and~(\ref{EQCD:1}) we see that terms linear in $p_i$
produce $-g_0' f^{a_1 a_2 a_3} V^{\mu_1 \mu_2 \mu_3}$,
the elementary 3-gluon vertex in the effective theory.
The structure $T_1^{\mu_1 \mu_2 \mu_3}$ gives $c^0_{g1}$,
the coefficient of the 3-gluon operator~(\ref{EQCD:O1})
in the effective Lagrangian~\cite{NSVZ:84}
\begin{equation}
c_{g1}(M) = - \frac{T_F}{90 M^2} \left( \frac{\alpha_s(M)}{4\pi}
+ \mathcal{O}(\alpha_s^2(M)) \right)\,.
\end{equation}
The structure $T_2^{\mu_1 \mu_2 \mu_3}$ reproduces~(\ref{EQCD:cg2}).
It is not difficult to extend this analysis to two loops using~(\ref{Qedland:V2}).

\subsection{Decoupling}
\label{S:DQCD}

Running of the full-theory coupling $\alpha_s^{(n_l+1)}(\mu)$
is governed by the $(n_l+1)$-flavour $\beta$-function;
running of the effective-theory coupling $\alpha_s^{(n_l)}(\mu)$
is governed by the $n_l$-flavour $\beta$-function;
their matching is given by
\begin{equation}
\alpha_s^{(n_l+1)}(\mu) = \zeta_\alpha(\mu) \alpha_s^{(n_l)}(\mu)\,,
\label{DQCD:alpha}
\end{equation}
with~\cite{LRV:95}
\begin{equation}
\zeta_\alpha(\mu_0)
= 1 + \left( \frac{13}{3} C_F - \frac{32}{9} C_A \right) T_F
\left(\frac{\alpha_s(\mu_0)}{4\pi}\right)^2 + \cdots
\label{DQCD:zeta}
\end{equation}
where $\mu_0$ is defined by~(\ref{Decoupling:mu0}).
Here the $C_F$ term can be obtained from the QED result~(\ref{charge:zeta2})
by inserting the obvious colour factors;
the $C_A$ term is more difficult to obtain.
The three-loop result has been obtained in~\cite{CKS:98},
and the four-loop one in~\cite{CKS:06,SS:06}%
\footnote{The result of~\cite{SS:06} contains one master integral which was not known analytically,
only numerically, with 37-digits precision.
An analytical expression for this integral has been published later~\cite{KKOV:06}.}.
The RG equation
\begin{equation}
\frac{d\log\zeta_\alpha(\mu)}{d\log\mu}
+ 2 \beta^{(n_l+1)}(\alpha_s^{(n_l+1)}(\mu))
- 2 \beta^{(n_l)}(\alpha_s^{(n_l)}(\mu)) = 0
\label{DQCD:RG}
\end{equation}
can be used to find $\zeta_\alpha(\mu)$ for $\mu\neq M$.

\begin{figure}[ht]
\begin{center}
\begin{picture}(100,80)
\put(50,40){\makebox(0,0){\includegraphics{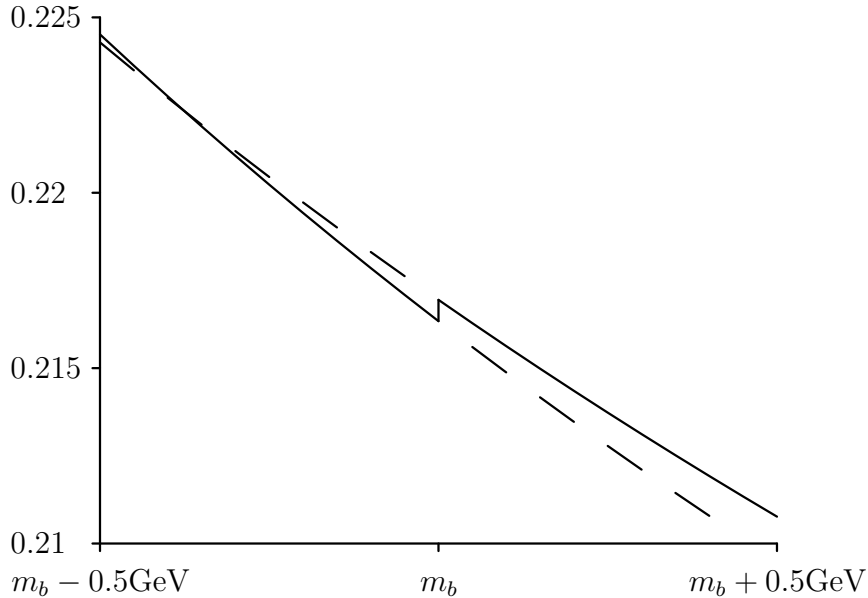}}}
\put(54,2){\makebox(0,0)[b]{$m_b$}}
\put(9,2){\makebox(0,0)[b]{$m_b-0.5\text{GeV}$}}
\put(99,2){\makebox(0,0)[b]{$m_b+0.5\text{GeV}$}}
\put(-3,9){\makebox(0,0)[l]{0.21}}
\put(-3,32.333333){\makebox(0,0)[l]{0.215}}
\put(-3,55.666667){\makebox(0,0)[l]{0.22}}
\put(-3,79){\makebox(0,0)[l]{0.225}}
\end{picture}
\end{center}
\caption{$\alpha_s^{(5)}(\mu)$ and $\alpha_s^{(4)}(\mu)$}
\label{F:as}
\end{figure}

The QCD running coupling $\alpha_s(\mu)$ not only runs when $\mu$ varies;
it also jumps when crossing heavy-flavour thresholds.
The behaviour of $\alpha_s(\mu)$ near $m_b$ is shown in Fig.~\ref{F:as}
(this figure has been obtained using the Mathematica package RunDec~\cite{CKS:00},
which takes into account four-loop $\beta$-functions and three-loop decoupling).
At $\mu>m_b$, the correct theory is the full 5-flavour QCD
($\alpha_s^{(5)}(\mu)$, the solid line);
at $\mu<m_b$, the correct theory is the effective low-energy 4-flavour QCD
($\alpha_s^{(4)}(\mu)$, the solid line);
the jump at $\mu=m_b$~(\ref{DQCD:zeta}) is shown.
Of course, both curves can be continued across $m_b$ (dashed lines),
and it is inessential at which particular $\mu\sim m_b$
we switch from one theory to the other one.
However, the on-shell mass $m_b^{\text{os}}$
(or any other mass which differs from it by $\mathcal{O}(\alpha_s)$,
such as, e.g., $\mu_0$)
is most convenient, because the jump is small, $\mathcal{O}(\alpha_s^3)$.
For, say, $\mu=2m_b$ or $\mu=m_b/2$ it would be $\mathcal{O}(\alpha_s^2)$.

Light-quark masses $m_i(\mu)$ also rum with $\mu$,
and also jump when crossing a heavy-quark threshold.
The QCD result
\begin{equation}
m^{(n_l+1)}(M) = m^{(n_l)}(M)
\left[1 - \frac{89}{18} C_F T_F
\left(\frac{\alpha_s(M)}{4\pi}\right)^2 + \cdots \right]
\label{DQCD:m}
\end{equation}
can be obtained from the QED one~(\ref{mass:zetam2})
by inserting the obvious colour factors.

\section{Conclusion}
\label{S:Conc}

In the past, only renormalizable theories were considered well-defined:
they contain a finite number of parameters,
which can be extracted from a finite number of experimental results
and used to predict an infinite number of other potential measurements.
Nonrenormalizable theories were rejected
because their renormalization at all orders in nonrenormalizable interactions
involve infinitely many parameters,
so that such a theory has no predictive power.
This principle is absolutely correct,
if we are impudent enough to pretend that our theory
describes the Nature up to arbitrarily high energies
(or arbitrarily small distances).

Our current point of view is more modest.
We accept the fact that our theories only describe the Nature
at sufficiently low energies (or sufficiently large distances).
They are effective low-energy theories.
Such theories contain all operators (allowed by the relevant symmetries)
in their Lagrangians.
They are necessarily nonrenormalizable.
This does not prevent us from obtaining definite predictions
at any fixed order in the expansion in $E/M$,
where $E$ is the characteristic energy
and $M$ is the scale of new physics.
Only if we are lucky and $M$ is many orders of magnitude larger
than the energies we are interested in,
we can neglect higher-dimensional operators in the Lagrangian
and work with a renormalizable theory.

Practically all physicists believe that the Standard Model
is also a low-energy effective theory.
But we don't know what is a more fundamental theory
whose low-energy approximation is the Standard Model.
Maybe, it is some supersymmetric theory (with broken supersymmetry);
maybe, it is not a field theory, but a theory of extended objects
(superstrings, branes);
maybe, this more fundamental theory lives in a higher-dimensional space;
or maybe it is something we cannot imagine at present.
The future will tell.

\textbf{Acknowledgements}.
I am grateful to K.\,G.~Chetyrkin and M.~Misiak for useful discussions.

\appendix
\section{Renormalization of bilinear currents}
\label{S:App}

Here we shall consider the currents
\begin{equation}
j_{n0} = \bar{\psi}_0 \gamma_{(n)} \tau \psi_0\,,
\label{App:j0}
\end{equation}
where
\begin{equation}
\gamma_{(n)} = \gamma^{[\mu_1}\cdots\gamma^{\mu_n]}\,,
\label{App:gamman}
\end{equation}
in QED with $n_f\ge2$ lepton flavours,
where $\tau$ is a flavour matrix with $\Tr\tau=0$
(flavour-singlet currents have some peculiarities,
because the fermion line emerging from the operator vertex
can return to the same vertex again;
we shall not discuss such currents).
The renormalized currents are related to the bare ones by
\begin{equation}
j_{n0} = Z_{jn}(\alpha(\mu)) j_n(\mu)\,,
\label{App:ren}
\end{equation}
where the $Z_{jn}$ are minimal renormalization constants.
The $\mu$ dependence of $j_n(\mu)$ is determined
by the renormalization-group equation
\begin{equation}
\left(\frac{d}{d\log\mu} + \gamma_{jn}(\alpha(\mu))\right) j_n(\mu) = 0\,,
\label{App:RG}
\end{equation}
where
\begin{equation}
\gamma_{jn} = \frac{d\log Z_{jn}}{d\log\mu}
\label{App:adim}
\end{equation}
is the anomalous dimension.

\begin{figure}[ht]
\begin{center}
\begin{picture}(56,20)
\put(28,11.5){\makebox(0,0){\includegraphics{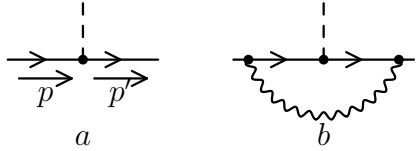}}}
\put(6,5){\makebox(0,0)[b]{$p$}}
\put(16,5){\makebox(0,0)[b]{$p'$}}
\put(11,0){\makebox(0,0)[b]{$a$}}
\put(43,0){\makebox(0,0)[b]{$b$}}
\end{picture}
\end{center}
\caption{Proper vertex of a bilinear current}
\label{F:jvert}
\end{figure}

Let the sum of one-particle-irreducible bare diagrams
with a current vertex $\gamma_{(n)}$,
an incoming fermion with momentum $p$,
and an outgoing fermion with momentum $p'$,
not including the external fermion propagators,
be the proper vertex $\Gamma_n(p,p')=\gamma_{(n)}+\Lambda_n(p,p')$
(Fig.~\ref{F:jvert}).
When the vertex is expressed via the renormalized quantities
$\alpha(\mu)$, $a(\mu)$,
it should become $Z_{\Gamma n}\Gamma_n^r(p,p')$,
where the renormalized vertex $\Gamma_n^r(p,p')$
is finite in the limit $\varepsilon\to0$.
When the proper vertex of the renormalized current $Z_{jn}^{-1}\Gamma_n(p,p')$
is multiplied by the two external-leg renormalization factors $Z_\psi^{1/2}$,
it should give a finite matrix element.
Therefore, $Z_{jn}=Z_q Z_{\Gamma n}$.
The UV divergences of $\Lambda_n(p,p')$
do not depend on the quark masses and the external momenta.
Therefore, we may assume that all fermions are massless,
and set $p=p'=0$.
An IR cut-off is then necessary in order to avoid
IR $1/\varepsilon$ terms.

In the one-loop approximation (Fig.~\ref{F:jvert}$b$),
\begin{equation*}
\Lambda_n(0,0) = -i e_0^2 \int \frac{d^d k}{(2\pi)^d} \frac{1}{(k^2)^3}
\left[\frac{\gamma^\mu\gamma^\nu\gamma_{(n)}\gamma_\nu\gamma_\mu}{d}
- (1-a_0)\gamma_{(n)}\right]\,,
\end{equation*}
where the averaging
$\rlap/k\gamma_{(n)}\rlap/k\to k^2\gamma^\alpha\Gamma\gamma_\alpha/d$ has been used.
Now we use an important identity
\begin{equation}
\gamma^\mu\gamma_{(n)}\gamma_\mu = (-1)^n (d-2n) \gamma_{(n)}\,.
\label{App:h}
\end{equation}
It can be understood as follows.
Let's take some specific component of $\gamma_{(n)}$.
Then $n$ values of the index $\mu$
which are present in this component of $\gamma_{(n)}$
contribute $-(-1)^n n$,
while the remaining $d-n$ values of $\mu$
contribute $(-1)^n (d-n)$.
Using the UV divergence
\begin{equation}
\left. \int \frac{d^d k}{(2\pi)^d}\; \frac{1}{(k^2)^2} \right|_{\text{UV}}
= \frac{i}{(4\pi)^2 \varepsilon}\,,
\label{App:UV}
\end{equation}
we obtain
\begin{equation}
Z_{\Gamma n} = 1 + \frac{\alpha}{4\pi\varepsilon}
\left[(n-2)^2-1+a\right]\,.
\label{App:ZG}
\end{equation}
The gauge dependence is cancelled by
\begin{equation}
Z_\psi = 1 - a \frac{\alpha}{4\pi\varepsilon}\,,
\label{App:Zpsi}
\end{equation}
and we arrive at
\begin{equation}
\gamma_{jn} = - 2 (n-1) (n-3) \frac{\alpha}{4\pi} + \cdots
\label{App:gamma1}
\end{equation}
Up to two loops~\cite{BG:95}
\begin{equation}
\begin{split}
\gamma_{jn} ={}& - 2 (n-1) (n-3) \frac{\alpha}{4\pi}
\left[1 + \frac{1}{2} \bigl(5(n-2)^2-19\bigr) \frac{\alpha}{4\pi} \right]\\
&{} - \frac{1}{3} (n-1) (n-15) \beta_0 \left(\frac{\alpha}{4\pi}\right)^2
+ \cdots
\end{split}
\label{App:gamma2}
\end{equation}
The three-loop result has been derived in~\cite{G:00}.

The vector current has $\gamma_{j1}=0$ to all orders.
For the vector current with a diagonal flavour matrix $\tau$,
the integral of its 0-th component is an integer ---
the difference between the numbers of light fermions and antifermions
weighted by the diagonal elements of $\tau$.
It does not depend on $\mu$.
The same holds for non-diagonal $\tau$ by flavour symmetry.
We can also see this explicitly.
Multiplying the Ward identity
\begin{equation}
\Gamma_1^\mu(p,p) = \frac{\partial S^{-1}(p)}{\partial p_\mu}
\label{App:Ward}
\end{equation}
by $Z_\psi$ transforms $S^{-1}$ into $S_r^{-1}$,
and hence makes the left-hand side finite.
Therefore, $Z_{j1}=Z_\psi Z_{\Gamma1}=1$.

The renormalization of the scalar current ($n=0$)
is closely related to that of the fermion mass $m$.
The \MS{} renormalized mass $m(\mu)$,
\begin{equation}
m_0 = Z_m(\alpha(\mu)) m(\mu)
\label{App:mass}
\end{equation}
(with a minimal renormalization constant $Z_m$),
is defined in such a way that the bare propagator
\begin{equation}
S(p) = \frac{1}{\rlap/p - m_0 - \Sigma(p)}
= \frac{1}{1-\Sigma_V(p^2)}\;
\frac{1}{\displaystyle\rlap/p -
\frac{1+\Sigma_S(p^2)}{1-\Sigma_V(p^2)} m_0}
\label{App:S}
\end{equation}
(where the fermion self-energy
$\Sigma(p)=\rlap/p \Sigma_V(p^2) + m_0 \Sigma_S(p^2)$),
when expressed via the renormalized quantities $\alpha(\mu)$,
$a(\mu)$, $m(\mu)$, is equal to $Z_\psi S_r(p)$,
where the renormalized propagator $S_r(p)$
is finite in the limit $\varepsilon\to0$.
In order to find $Z_m$,
it is sufficient to consider $|p^2|\gg m^2$
and retain only terms linear in $m$.
Differentiating diagrams for $\Sigma$ in $m_0$, we get
\begin{equation}
1 + \left.\Sigma_S(p^2)\right|_{m_0=0} = \Gamma_0(p,p)\,.
\label{App:Sigma}
\end{equation}
The renormalization constant $Z_m$ is defined by the condition
that $Z_q Z_m (1+\Sigma_S)$ is finite.
Therefore,
\begin{equation}
Z_m = Z_{j0}^{-1}\,.
\label{App:Zm}
\end{equation}
In other words,
$m(\mu)\left[\bar{\psi}\tau\psi\right]_\mu=m_0\bar{\psi}_0\tau\psi_0$
is not renormalized.
The anomalous dimension of the mass is
\begin{equation}
\gamma_m = -\gamma_{j0} =
6 \frac{\alpha}{4\pi}
+ \left(3 - \frac{20}{3} n_f\right)
\left(\frac{\alpha}{4\pi}\right)^2 + \cdots
\label{App:gammam}
\end{equation}

It is not possible to introduce an anticommuting $\gamma_5$
satisfying $\gamma_5^{\text{AC}}\gamma^\mu+\gamma^\mu\gamma_5^{\text{AC}}=0$
in $d$ dimensions.
A way out was proposed by 't~Hooft and Veltman.
Let us split our $d$-dimensional space--time
into a 4-dimensional subspace
and the orthogonal $(d-4)$-dimensional subspace,
and define $\gamma_5^{\text{HV}}=-i\gamma^0 \gamma^1 \gamma^2 \gamma^3
=(i/4!)\varepsilon_{\alpha\beta\gamma\delta}
\gamma^\alpha \gamma^\beta \gamma^\gamma \gamma^\delta$,
where the tensor $\varepsilon_{\alpha\beta\gamma\delta}$
lives in the 4-dimensional subspace.
This $\gamma_5^{\text{HV}}$ anticommutes with $\gamma^\mu$
from the 4-dimensional subspace and commutes with those
from the $(d-4)$-dimensional one.
This consistent definition leads to the anomaly
of the flavour-singlet axial current.
But the fermion line cannot be closed in matrix elements
of flavour-nonsinglet currents.
Therefore, traces with a single $\gamma_5$,
which can lead to an anomaly, never appear,
and one may use a naively anticommuting $\gamma_5^{\text{AC}}$
without encountering contradictions.
The pseudoscalar currents
$j_{\text{AC}}(\mu)=
Z_{j0}^{-1}(\alpha(\mu)) \bar{\psi}_0\gamma_5^{\text{AC}}\tau\psi_0$
and
$j_{\text{HV}}(\mu)=
Z_{j4}^{-1}(\alpha(\mu)) \bar{\psi}_0\gamma_5^{\text{HV}}\tau\psi_0$
are related to each other by a finite renormalization:
\begin{equation}
j_{\text{AC}}(\mu) = Z_P(\alpha(\mu)) j_{\text{HV}}(\mu)\,,\quad
Z_P(\alpha) = 1 + z_{P1} \frac{\alpha}{4\pi}
+ z_{P2} \left(\frac{\alpha}{4\pi}\right)^2 + \cdots
\label{App:ZPdef}
\end{equation}
Similarly, the axial currents are related by
\begin{equation}
j_{\text{AC}}^\mu(\mu) = Z_A(\alpha(\mu)) j_{\text{HV}}^\mu(\mu)\,.
\label{App:ZAdef}
\end{equation}
This is clearly discussed in~\cite{L:93},
where references to earlier papers can be found.

The finite renormalization constants $Z_{P,A}$
can be found by calculating matrix elements of
the operator equalities~(\ref{App:ZPdef}), (\ref{App:ZAdef}).
The easiest choice is the matrix element from an initial fermion
to a final one with momenta $p$, $p$ (or $p$, 0).
At one loop (but not beyond) $Z_{j4}=Z_{j0}$, $Z_{j3}=Z_{j1}=1$;
calculating the proper vertices and equating the matrix elements,
we can easily get
\begin{equation}
Z_P(\alpha) = 1 - 8 \frac{\alpha}{4\pi} + \cdots\,,\quad
Z_A(\alpha) = 1 - 4 \frac{\alpha}{4\pi} + \cdots
\label{App:ZPA1}
\end{equation}
They can be also obtained from the anomalous dimensions of the currents.
Differentiating~(\ref{App:ZPdef}) and~(\ref{App:ZAdef}), we have
\begin{equation}
\begin{split}
\frac{d\log Z_P(\alpha)}{d\log\alpha} &{}=
\frac{\gamma_{j0}(\alpha)-\gamma_{j4}(\alpha)}{2\beta(\alpha)}\,,\\
\frac{d\log Z_A(\alpha)}{d\log\alpha} &{}=
\frac{\gamma_{j1}(\alpha)-\gamma_{j3}(\alpha)}{2\beta(\alpha)}\,,
\quad\text{where {}}\gamma_{j1}=0\,.
\end{split}
\label{App:RGPA}
\end{equation}
The one-loop results~(\ref{App:ZPA1}) can be reproduced
using the two-loop anomalous dimensions~(\ref{App:gamma2}).
Now we see the reason why the last term in~(\ref{App:gamma2}),
which is not symmetric with respect to $n\to4-n$,
is proportional to $\beta_0$.
As $Z_{P,A}$ can be obtained solely from the anomalous dimensions,
they are determined by the UV behaviour of the matrix elements,
and cannot depend on masses and external momenta.

The two-loop results
\begin{equation}
\begin{split}
Z_P(\alpha) &{}= 1 - 8 \frac{\alpha}{4\pi}
+ \frac{8}{9} n_f \left(\frac{\alpha}{4\pi}\right)^2 + \cdots\,,\\
Z_A(\alpha) &{}= 1 - 4 \frac{\alpha}{4\pi}
+ \frac{2}{9} \left( 99 + 2 n_f \right)
\left(\frac{\alpha}{4\pi}\right)^2 + \cdots
\end{split}
\label{App:ZPA2}
\end{equation}
can be obtained either from two-loop matrix elements
or from the three-loop anomalous dimensions.
The three-loop results have been calculated~\cite{LV:91,L:93}
from the matrix elements only.

Why have we not discussed a similar relation
$j_{\text{AC}}^{\mu\nu}(\mu)=Z_T(\alpha(\mu))j_{\text{HV}}^{\mu\nu}(\mu)$
between the tensor currents $j_{0\text{AC}}^{\mu\nu}
=\bar{\psi}_0\gamma_5^{\text{AC}}\sigma^{\mu\nu}\tau\psi_0$
and $j_{0\text{HV}}^{\mu\nu}
=\bar{\psi}_0\gamma_5^{\text{HV}}\sigma^{\mu\nu}\tau\psi_0$?
The current $j_{\text{AC}}^{\mu\nu}$
has the same anomalous dimension $\gamma_{j2}$
as the current $j_2^{\mu\nu}$ without $\gamma_5^{\text{AC}}$.
Multiplication of $\sigma^{\mu\nu}$ by $\gamma_5^{\text{HV}}$
is merely a space--time transformation,
e.g., $\gamma_5^{\text{HV}}\sigma^{01}=-i\sigma^{23}$,
and hence $j_{\text{HV}}^{\mu\nu}$ has the same anomalous dimension, too.
Therefore, $Z_T(\alpha(\mu))$ cannot depend on $\mu$,
and we conclude that $Z_T(\alpha)=1$.

\end{document}